\documentclass[a4paper,11pt]{article}
\pdfoutput=1 

\usepackage{jheppub} 

\usepackage[T1]{fontenc} 

\usepackage{amssymb}
\usepackage{graphicx}

\usepackage{xcolor}

\title{\boldmath Multi-kinks and composite oscillons in a commensurable and non degenerate double sine-Gordon model}

\author[a]{Jonathan Lozano-Mayo,}
\author[b,1]{Manuel Torres-Labansat,\note{Corresponding author.}}

\affiliation[a]{Weinberg Institute for Theoretical Physics, Department of Physics, The University of Texas at Austin, Austin,\\ TX 78712, USA}
\affiliation[b]{Instituto de F\'{\i}sica,
Universidad Nacional Aut\'onoma de M\'exico,\\
Apartado Postal 20-364, Ciudad de M\'exico  01000, M\'exico}

\emailAdd{Jonathanloz@utexas.edu}
\emailAdd{torres@fisica.unam.mx}

\abstract{ In this paper, we introduce a commensurable and non-degenerate double  sine-Gordon model, in which a partial breaking of vacuum degeneracy provides a mechanism for the emergence of static multi-kinks. These multi-kinks  $K_n$ are stable field configurations with internal structure, consisting of $n$ localized energy packets with well-defined separations. The properties of the multi-kinks are thoroughly analyzed, including the novel phenomenology that arises during their collisions. In particular, we observe the emergence of long-lived composite oscillons that reflect the original structure of the multi-kinks. The sub-kink's positions and their vibration modes provide collective coordinates that are used to construct a phenomenological  model, which offers a good qualitative explanation of the observed oscillon properties.     Linear radiation effects are consistently incorporated into the model, revealing their important role in the synchronization of the oscillon's vibrational components. In contrast, oscillon decay and the sudden annihilations observed in some collisions are driven by a novel staccato-like mechanism, whereby repeated crossings of the mass threshold by the dominant frequency generate a sequence of radiation bursts.     }


\arxivnumber{2407.09682}

\begin{document} 
\maketitle
\flushbottom

\section{Introduction}
 \label{sec:intro}

Scalar field theories provide a versatile framework for modeling a wide variety of physical phenomena. In particular, they offer a natural setting for spontaneous symmetry breaking, a mechanism that underlies both phase transitions and the generation of particle masses through the Higgs mechanism. In the standard picture, spontaneous symmetry breaking arises as a control parameter of the potential is varied, producing multiple degenerate vacua. In some models, tuning an additional parameter can partially lift this degeneracy. The present work is devoted to exploring the dynamical consequences of such situations.

A distinctive feature of nonlinear scalar field theories is their ability to support solitons: spatially localized, stable, finite-energy configurations having a conserved topological charge \cite{Manton2004,Vachaspati2006,Weinberg2012}. Even the simplest one-dimensional example, the $\phi^4$ model with two degenerate minima, exhibits kink solutions whose dynamics remain an active subject of study \cite{Dashen1974,Campbell1983,Kevrekidis2019}. Of particular interest are kink-antikink collisions. At low impact velocities these typically produce long-lived bound states, or bions \cite{Makhankov1978}. At higher velocities, collisions may lead to inelastic reflections in which the solitons escape to infinity after undergoing several bounces. The resulting outgoing-incoming velocity diagrams display resonance windows characteristic of chaotic scattering \cite{Moshir1981,Campbell1983,Anninos1991,Simas2016}. Much of this phenomenology can be qualitatively captured by reducing the full field theory to a small number of collective coordinates, most commonly the kink position and its internal (shape) mode. In this description, multi-bounce events arise from reversible energy exchange between translational and vibrational modes \cite{Sugiyama1979,Takyi2016,Manton21A,Manton21B}.

Another paradigmatic example is the sine-Gordon (SG) model \cite{Scott2003,Cuevas-Maraver2014}, an exactly integrable system whose analytical  solutions can be constructed using the inverse scattering method \cite{Ablowitz1973,Zakharov1973}. Besides static kinks, the SG model admits breathers and multi-soliton solutions, in which scattering occurs without radiation and solitons emerge unchanged apart from a phase shift. Many generalizations of sine-Gordon theory have been explored extensions that break exact integrability but retain some SG-like characteristics \cite{Campbell1986}.

Models featuring multiple degenerate vacua have also attracted significant attention. The $\phi^6$ model is one of the most studied examples, with extensive analyses of its kink-antikink dynamics and resonance structure \cite{Dorey2011,Demirkaya2017,Moradi2017,Gani2020}. Similar models are relevant for describing both first- and second-order phase transitions \cite{Khare2014,Zurek2024}, as well as long-range and non-exponential inter-kink forces \cite{Manton2019,Christov2019,Belendryasova2019,Christov2021}. Some of the aforementioned models support multi-kink solutions with internal structure \cite{Christ1975,Brito2014,Demirkaya2017,Lozano2021,Dorey2023,Halcrow2024}.

Nonlinear scalar theories also host oscillons: spatially localized, quasi-periodic, exceptionally long-lived configurations despite having no conserved topological charge. Beyond their intrinsic mathematical interest, oscillons appear in diverse  high energy physics and cosmological scenarios, including post-inflationary reheating and phase-transition dynamics, see, e.g., \cite{Gleiser1994,Kolb1994,Graham2007,Amin2010,Rompineve2021}.

In this work we introduce a commensurable and non-degenerate double $K_n$ sine-Gordon model, which naturally brings together many of the themes mentioned above. In this model, the usual sine-Gordon potential is modulated by a second periodic term, effectively extending the periodicity from $\phi \to \phi + 2\pi$ to $\phi \to \phi + 2\pi n$.
This structure produces $n$ inequivalent kink solutions $k_i$, $i=1,\ldots,n$. Introducing a degeneracy-breaking parameter $\alpha$ lifts the degeneracy between internal vacua while preserving the external ones, causing the individual kinks to merge into a single composite multi-kink configuration denoted as $K_n$.

Exact analytic expressions for $K_n$ are not available, but we derive accurate approximate profiles that closely match the numerical solutions. These enable a detailed study of the multi-kink's structure, including inter-kink forces, equilibrium separations, topological charge, size, and mass.

We also investigate multi-kink collisions. By varying the degeneracy-breaking parameter $\alpha$ and the initial impact velocity, we uncover a diverse range of outcomes:  inelastic reflections, annihilation processes, and the formation of long-lived composite oscillons.

To interpret these results, we develop an effective collective coordinate model based on the positions $x_i(t)$ of the internal kinks and their vibrational modes. Crucially, we show that including radiation effects in the effective theory is essential to reproduce the synchronization phenomena observed in the oscillon dynamics of the full field theory. Finally, we present results that suggest that  
oscillons decay  through  a staccato-like  radiation mechanism.

The paper is organised as follows. Section~\ref{GSGM} introduces the commensurable and non-degenerate double sine-Gordon model.
Section~\ref{SECCCM} presents the derivation of the collective coordinate model.
Section~\ref{MKAMKC} analyzes collisions of the $K_n$ multi-kinks with their corresponding    ${\bar K}_n$ anti-multi-kink.
Final remarks are given in section~\ref{conclu}.

\section{The commensurable and non degenerate double sine-Gordon model}\label{GSGM}

\subsection{General considerations}\label{Gcons}
 
 The Lagrangian density  for the  scalar field   $\phi(t,z)$   is  

\begin{equation}
 \label{Lagrang}
  \mathcal{L}=\frac{1}{2} \left(\frac{\partial \phi}{\partial t} \right)^2 -   \frac{1}{2} \left(\frac{\partial \phi}{\partial z}\right)^2 -  V_{n,\alpha} .
\end{equation}
Thus, the energy functional corresponding to this Lagrangian is given by the following expression 
\begin{equation}
\label{Enertot}
     E\left[\phi \right] = \int_{-\infty}^{\infty} {\cal E}(t,z) dz =\int_{-\infty}^{\infty}  \left(\frac{1}{2}\left(\frac{\partial \phi}{\partial  t}\right)^2+  \frac{1}{2}\left(\frac{\partial \phi}{\partial z}\right)^2+  V_{n,\alpha}  \right) dz ,
\end{equation}  
whereas  the field equation of motion is given as  
\begin{equation}
\label{eqphi}
     \frac{\partial^2\phi}{\partial t^2}-\frac{\partial^2\phi}{\partial z^2}=-\frac{d V_{n,\alpha}}{d \phi}.
\end{equation}  
Focusing on static configurations, the previous equation becomes  
 \begin{equation}
\label{eqphi1}
   \frac{d^2\phi}{d z^2}= \frac{d V_{n,\alpha}}{d \phi},
   \end{equation}
  multiplying this equation  by $d \phi /dz$  and integrating yields the first order Bogomol'nyi equation \cite{Bogomolny1976} 
\begin{equation}
\label{Bogo}
  \frac{1}{2}  \left(  \frac{d \phi}{d z}  \right)^2=  V_{n,\alpha}(\phi) , 
   \end{equation}
 where the integration constant is selected as  zero, so the  kink energy  obtained from   eq.~($\ref{Enertot}$) is  finite. We find convenient to define the superpotential $W_{n,\alpha}(\phi)$ \cite{Boya1989,Manton2023} as
\begin{equation}
 \label{Suppot}
 W_{n,\alpha}(\phi) =    \int^ \phi \sqrt{2 \, V_{n,\alpha}(\phi)}  \, d \phi , 
   \end{equation}
 this expression together with the Bogomol'nyi   eq.~(\ref{Bogo})  imply the relation $ (d \phi/ dz) = (d  W_{n,\alpha} / d \phi)$. 
 
\subsection{The  potential}\label{Vpot}
\begin{figure}
\begin{center}
\includegraphics[width=10.0cm]{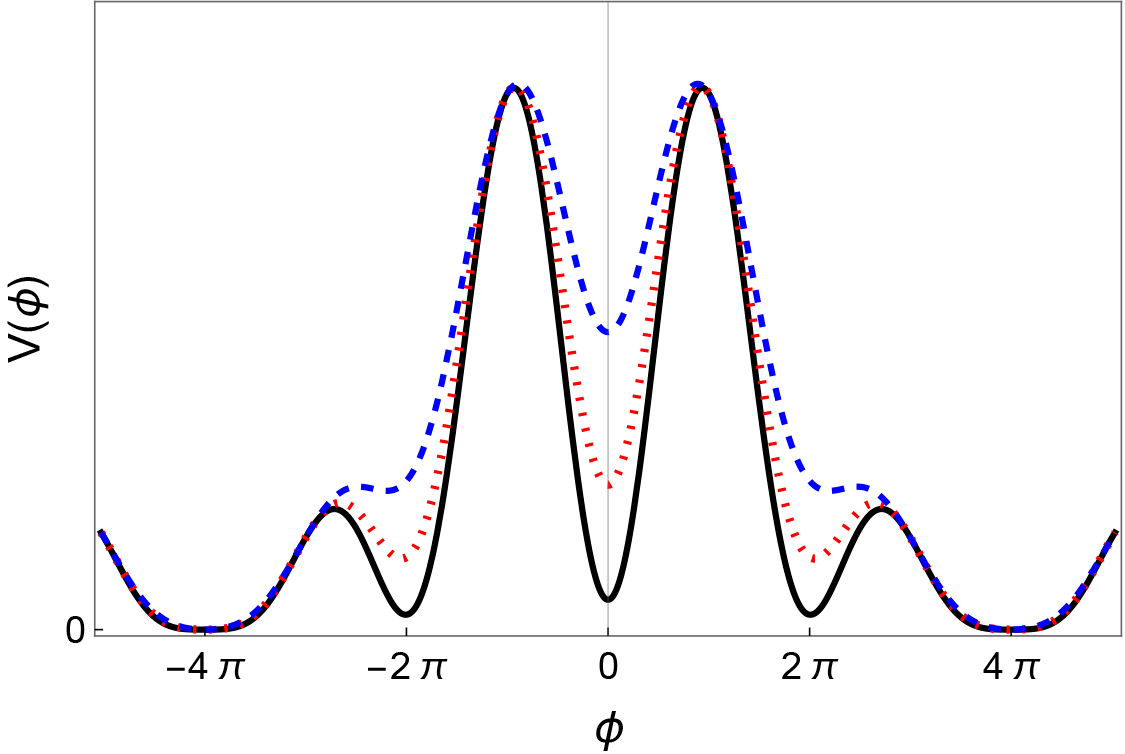}
\caption{Potential  $V_{n,\alpha} (\phi) $, the parameters are selected as: $n= 4$, with: $ \alpha=0.1 $ (black-continuous line), $ \alpha=0.6 $ (red-dotted line) and $ \alpha=1.5 $ (blue-dashed line).  }
\label{Figupot}
\end{center}
\end{figure}

In the  proposed model   the usual Sine-Gordon  potential $   (   1 - \cos \phi)$  is  modulated by a second Sine-Gordon-like term $\left( 1 + \cos \frac{\phi}{n}  \right)$ and  shifted by the degeneracy-breaking parameter  $ \alpha$. As a result, the potential 

\begin{equation}
\label{Pot2SG}
V_{n,\alpha} \left(  \phi \right)  =      \frac{1}{2(1 + \frac{1}{2} \vert  \alpha \vert ) } \left(   1 - \cos \phi +   \alpha \right) \left( 1 + \cos \frac{\phi}{n}      \right), 
 \end{equation}
takes the form shown in   figure~\ref{Figupot}.
 For future reference, it is convenient to split the potential as $ V_{n,\alpha} \left(  \phi \right) = V_{n,0} \left(  \phi \right) + \alpha \Delta_{n,\alpha} \left(  \phi \right)$, with
 \begin{equation}
\label{Deltaphi}
\Delta_{n,\alpha} \left(  \phi \right) = \frac{2 }{2 + \alpha }\left( \cos\frac{\phi}{2 } \, \, \, \cos\frac{\phi}{2n} \right)^2   . 
 \end{equation}
 The potential reduces to the usual Sine-Gordon  form when $\alpha = 0 $ and $ n \to \infty$. In the opposite limit, as $\alpha  \to \infty$,  it also approaches a SG potential under the field redefinition 
 $\phi \to \frac{\phi}{n}+\pi$. The potential $V_{n,\alpha}$  is invariant under the   transformations $\phi \to -\phi $ and   $\phi  \to   \phi + 2 \pi n$. Unless otherwise specified, we focus on solutions within the sector $\phi  \in   \{ - \pi n ,  \pi n  \}$ sector.  For simplicity,  we assume that  $n$ is an even integer, similar results hold for  odd $n$.

The $n+1$ potential minima can be  determined numerically, a useful analytical approximation can be obtained in the small-$\alpha$ regime
\begin{equation}
\label{vi}
v_{ i} \approx    \ 2 i   \pi  + \frac{\alpha}{2n} \frac{\sin \left( \frac{2 \pi i}{n} \right)}{\cos^2 \left( \frac{\pi i}{n} \right) },
   \,\,\,\,\,\,  i=  \pm \left( 0,1,2,...., \frac{n}{2}-1 \right) , \qquad v_{ \pm \frac{n}{2}} =\pm n \pi . 
 \end{equation}
 Associated with each potential minimum, there are perturbative scalar excitations whose masses are determined by  $m_{ i}^2 = \frac{ d^2 V_{n,\alpha} \left(  \phi \right) }{d^2 \phi } { \Big \vert}_{\phi=  v_{i}} $, which, for small $\alpha$, are approximately given by:
 \begin{equation}
\label{msc}
m_{ i } \approx  
 \sqrt{  \cos^2 \left( \frac{i  \, \pi}{n} \right) -  \frac{\alpha}{4n^2} 
   \left(  6+ n^2 +( n^2-4 )\cos \left( \frac{2\pi i}{n} \right) \right)} \,, 
   \qquad m_{\pm \frac{n}{2}} = \sqrt{ \frac{\alpha}{2n^2} }.
 \end{equation} 
  
For $ \alpha = 0$, the model admits $n$   kink solutions $ k_i$. As  $\alpha$  increases within the $0  < \alpha \le \alpha_1$ interval, these $n$ kinks  merge into a single multi-kink configuration $K_n$, characterized by well-defined localized energy lumps. The value $\alpha_1$  marks the onset of significant overlap among sub-kinks, will be determined later in  eq.~($\ref{alpha1}$).
In the subsequent interval $\alpha_1  < \alpha \le \alpha_2$ ,  the distinction between individual sub-kinks becomes less pronounced. The  $\alpha_2$   threshold  corresponds to the value at which the internal local minima of the potential start disappearing. This critical value can be estimated by requiring that the  mass $m_{\frac{n}{2}-1}$ of the internal scalar excitation, given in  eq.~($\ref{msc}$), vanish. This leads to
\begin{equation}
\label{alphacr}
\alpha_{2} =       \frac{ 4 n^2   \sin^2 ( \frac{ \pi}{n} )}  { 6+ n^2  -(n^2 -4) \cos ( \frac{ 2\pi}{n} )}.
\end{equation}
For $\alpha > \alpha_2 $, the field configuration converges to a single soliton, as the internal substructure associated to the  multi-kinks is gradually erased.
Notice  that $\alpha_2 $ has a weak dependence on $n$,   $\alpha_2  \in \{1.6,1.33 \} $ for $n  \in \{2, \infty \} $.
 Although stable soliton solutions are not expected for negative values of $\alpha$, the model admits the existence of saddle-point solutions that interpolate between false and true vacua. These unstable, energy-maximizing configurations known as sphalerons \cite{Manton2023,Izquierdo2023} represent transitions over the potential barrier and will be explored in future work.

 \subsection{$n$ kinks $k_i$  in the  $\alpha=0$  degenerate regime}\label{nkisec}
 
 For  $\alpha=0$  the potential minima are located at $v_{i} =    2 i \pi$  with $i=\pm (0,1,2,...., \frac{n}{2})$, so the model  admits  $n$ kink solutions   $ \phi_{k_i}   \in   \{ v_i ,v_{i+1}   \}$, $i=-\frac{n}{2},...,0,...., \frac{n}{2}$, each carrying  the same topological charge  $q_{k_i} = 
\frac{1}{2 \pi}(v_{i+1}-v_i)=1$, along with their  corresponding antikinks   ${\bar k}_i $, with charge
 $q_{\bar k_i} =-1$. 
The location of  the potential maxima $ \Omega_i $ deviate  from those of the standard sine-Gordon model, where
 $ \Omega_{ i} = \pm (2i - 1 ) \pi  $. In this case, their positions are determined by the solutions to the transcendental equation $n \cot (\frac{\phi}{2}) - \tan (\frac{\phi}{2n}) =0$.  Notice that for $\alpha=0$, the masses $m_{\pm\frac{n}{2}}$ vanishes.

The superpotential,  given  in  eq.~($\ref{Suppot}$) takes the form,
\begin{equation}
\label{SupPot}
W[\phi] =-\frac{4n}{n^2 -1 }  \left(  n  \cos  \frac{\phi}{2 } \cos  \frac{\phi}{2n } + \sin \frac{\phi}{2 } \sin  \frac{\phi}{2n }  \right),
 \end{equation}
from which the kink masses are computed as:
 \begin{equation}
\label{MkiSG}
M_{k_{i}} =W[v_{i}] - W[v_{i-1} ] =\frac{4n^2}{n^2 -1 }  \left(  \cos  \frac{v_{i}}{2n }  + \cos  \frac{v_{i-1} }{2n }   \right),   \qquad i= \pm \left( 1,2,...., \frac{n}{2} \right). 
 \end{equation}
  It is interesting to observe that, using   eqs.~(\ref{msc},\ref{MkiSG}),  the masses  $M_{k_i}$ are related to the scalar field masses evaluated at the neighboring minima $v_i$ and   $v_{i-1}$ through the following expression:
$M_{k_{i}} = \frac{4n^2}{n^2 -1 }  \left(  m_{i} + m_{i-1}  \right)$.

 The  soliton $\phi_{k_i}(z)$  configurations correspond to the solutions of the equation  
 \begin{equation}
\label{BESG}
\frac{d \phi}{d z}  = 2 \sin  \frac{ \phi }{2} \cos  \frac{\phi }{2n}.
 \end{equation}
For $n=2$,  eq.~(\ref{BESG}) admits the implicit analytical solution $ \ln \left[  \tan( \frac{ \phi}{8}) \right] + \sec( \frac{ \phi}{4} )= z$. For any other value of $n$, the equation must be solved numerically. However, a very good analytical approximation can be derived by noting that, within each kink $ \phi_{k_i}  \in   \{ v_{i-1} ,v_{i}   \}$ sector, the function $\cos  \frac{\phi }{2n}$ varies  slowly  compared to   $\sin  \frac{ \phi }{2}. $ This motivates an approach we refer to as the smooth-modulation approximation, which becomes increasingly accurate as $n$, grows. Under this approximation, if the  $\cos  \frac{\phi }{2n}$ term  is evaluated at the boundary  points $   \phi  \sim    v_{i-1}$ or  $  \phi  \sim v_{i}   $,   the solutions of   eq.~(\ref{BESG}) coincide with the well known SG field  solution with the asymptotic behavior $ \phi (z) \, \sim v_{i-1} + const \, e^{m_{i-1} z }$  or $ \phi (z) \, \sim v_{i}  - const \,e^{-m_{i} z }$ as $ z \to \mp \infty$. 
 These results hold for all kinks except those near the absolute minima, where according to  eq.~(\ref{msc})  the scalar mass $m_{\pm n/2}$  vanishes. In these special cases, the correct asymptotic behavior becomes
  $ \phi (z) \, \sim  \mp n \pi  \pm   \,\frac{4n}{z}  $. This analysis suggests an ansatz for the kink solution $ \phi_{k_i} (z) \in   \{ v_{i-1} ,v_{i}   \}$, that reproduces the correct asymptotic forms in each sector, and is given by 
  \begin{align}
\label{phiAn} 
 \phi_{k_i} (z) \,  = v_{i-1} \,   +  & \, 4  \,  \tan^{-1}   G_i \left[  z  \right],   \qquad i=1......, \frac{n}{2},  \\ \nonumber 
   G_i \left[  z  \right] = \frac{1}{ a_i e^{-m_{i-1} z}  + b_i e^{-m_{i} z} }, \,\,\, i< \frac{n}{2},  &\qquad  G_{\frac{n}{2}} \left[  z  \right] = \ln
  \left[  1 + \frac{1}{ c_n\, e^{- m_{\frac{n}{2}-1} z}  + d_n \, e^{- z/n} }  \right].
\end{align}
The solutions  for negative values of  $i$ are obtained   via spatial and field reflection symmetry, using the relation  $\phi_{k_{(-i)}} (z) = -\phi_{k_{i}} (-z)$, $i=1......, \frac{n}{2}$. Given that $m_{i-1} > m_{i}$, the  $ \phi_{k_i} (z)$ ansatz correctly reproduces the asymptotic  exponential tails determined by the scalar masses $m_{i-1}$ (for $ z \to - \infty$ ) and $m_{i}$ (for $ z \to + \infty$ ).  In the  $i= \frac{n}{2}$ case, the function $G_{\frac{n}{2}} \left[  z  \right]$ 
  captures  both the  exponential $ \phi (z) \, \sim  (n-2)  \pi  + \frac{4}{c_n}  \,e^{-m_{\frac{n}{2} -1} z }$ and long range $ \phi (z) \, \sim n \pi - \frac{4 n}{z}  $ asymptotic  tail behavior.  The coefficients $a_i$, $b_i$, $c_n$ and $d_n$  are determined by enforcing two conditions: $(i)$ at the kink center the field value matches  the location of the potential maxima, $i.e.$ $\phi_{k_i} (0) = \Omega_{i}$,  and $(ii)$ the field derivative at this point satisfies   eq.~(\ref{BESG}).   Using these conditions, the explicit expressions for the coefficients are derived as 
  \begin{equation} 
  \label{phiAn2}
  a_i = \frac{\gamma_i  \mu_i- m_{i} }{\gamma_i (m_{i-1} -m_{i} )} \qquad   b_i= \frac{m_{i-1} -\gamma_i  \mu_i}{\gamma_i (m_{i-1} -m_{i} )},
  \qquad i=1......, \frac{n}{2}-1,
   \end{equation} 
where $  \gamma_i = \tan \frac{\Omega_i -v_{i-1}}{4} $ and
$  \mu_i  =  \sin \frac{ \Omega_i }{2} \cos \frac{ \Omega_i}{ 2n } /\left(2 \sin^2 \frac{(\Omega_i - v_{i-1})}{4} \right)$, whereas 
   
    \begin{equation} 
  c_n = \frac{n  \nu_n- \kappa_n  }{n \sin \frac{\pi}{n}-1} \qquad   d_n= \frac{n (\kappa_n \sin \frac{\pi}{n} - \nu_n )}{n \sin \frac{\pi}{n}-1},
   \end{equation} 
   with  $  \nu_n  =  \frac{ \kappa_n}{4} (\kappa_n  +1) (1+  \gamma_n^2)\sin \left( \frac{ \Omega_{n/2} }{2}  \right)   \cos   \left(\frac{ \Omega_{n/2}}{ 2n }  \right) $,   $\kappa_n=(e^{\gamma_n}-1)^{-1}$, and $  \gamma_n = \tan \frac{\Omega_{n/2} -(n-2)\pi}{4} $.
 
\begin{figure}
 \begin{center}
\includegraphics[width=16.0cm]{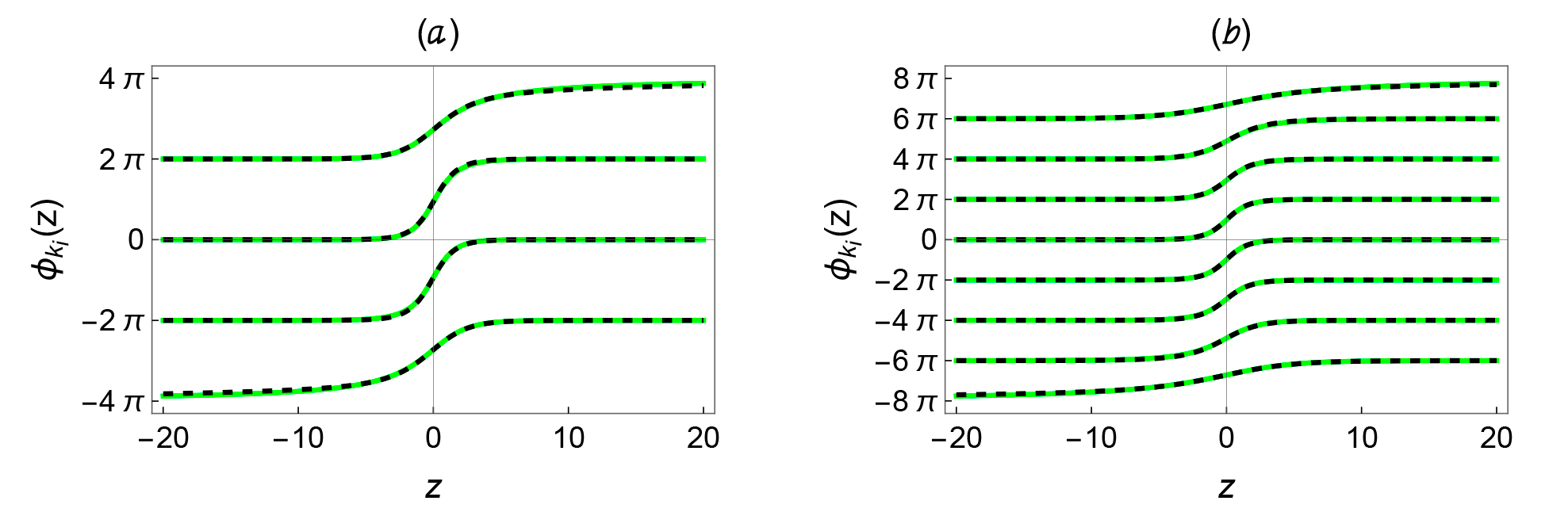}
\caption{Kink  field configurations $\phi_{k_i } (z)$  obtained from the numerical solution  of  eq.~(\ref{BESG})   (continuous-green lines)  and the  approximated ansatz    eq.~(\ref{phiAn})  (dashed-black lines); for the  (a)  $n=4$  and (b)  $n=8$ cases.
  }
\label{Figphii}
\end{center}
\end{figure}

 Figure \ref{Figphii}  presents the  $\phi_{k_i} (z) $ kinks profiles obtained  from  the numerical solutions of    eq.~(\ref{BESG}) as compared with the approximated ansatz  eq.~(\ref{phiAn}). The results clearly show that the smooth-modulation approximation performs remarkably well, particularly for $n  >2$. 

  \subsection{Multi-kink $K_n$  in the  $\alpha  >0$  non-degenerate regime}\label{Knsec}
  
  In this section, we analyze the properties of the $K_n$ multi-kink that emerges in the $\alpha > 0$ phase. Although the minima $v_i$ are no longer located at integer multiples of $2\pi$, the topological charge $Q_n$ remains straightforward to compute. This is because the values at the absolute minima, $v_{\pm n/2} = \pm n\pi$, are not affected. Consequently, the multi-kink interpolates between $\phi_{z \to -\infty} = v_{-n/2}$ and $\phi_{z \to \infty} = v_{n/2}$, yielding a total topological charge   $Q_n =  \frac{1}{2 \pi}(v_{n/2}- v_{-n/2})=n$, which corresponds to the sum of the topological charges of the $n$ constituent sub-kinks $Q_n =\sum_{i}  q_{k_i}=n$.
  
  \subsubsection{Multi-kink mass $ M_{K_{(n,\alpha)}} $ }

\begin{figure}
 \begin{center}
\includegraphics[width=16.0cm,height=6.0cm]{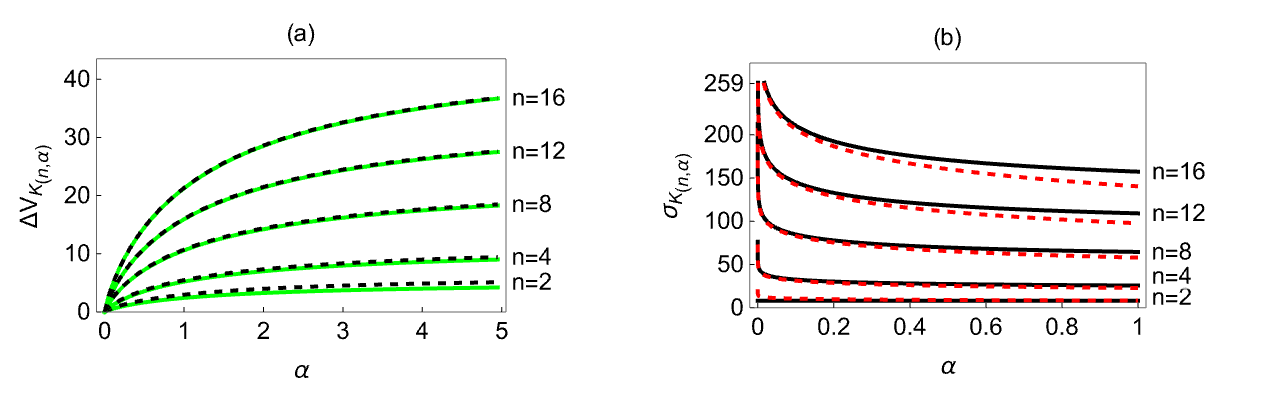}
\caption{  $K_n$  multi-kink: $(a)$  Binding energy $\Delta V_{K_{(n,\alpha)}}$    as a function of  $\alpha$, obtained from the numerical integration of the second term of   eq.~(\ref{MKSM1})   (continuous-green lines)  and the analytical expression in  eq.~(\ref{Vint}) (dashed-black lines). $(b)$ Size $ \sigma_{K_{(n,\alpha)}}$   obtained from  eq.~(\ref{sigmaK}) (continuous-black lines), the dashed lines correspond to the approximated expression in  eq.~(\ref{sigmaK2}). 
  }
\label{FigBen}
\end{center}
\end{figure}

Other  multi-kink  properties  can be analyzed using the smooth-modulation approximation. To compute the total mass  $ M_{K_{(n,\alpha)}} $, we decompose it into two contributions: the sum of the internal kink masses, $M_{K(n,0)}$, and the $\alpha$-dependent binding interaction energy,  $\Delta V_{K(n,\alpha)}$. These correspond to the first and second terms, respectively, in the following expression:
 \begin{equation}
  \label{MKSM1}
M_{K(n,\alpha)}\, =   \int_{- n  \pi  }^{  n  \pi }    \sqrt{2 V_{n, 0} (\phi) } d\phi +    \int_{- n  \pi  }^{  n  \pi }   \left(  \sqrt{2 V_{n, \alpha} (\phi) } - \sqrt{2 V_{n, 0} (\phi) }  \right) d\phi \, .
 \end{equation} 
The first term is the sum of the internal kink masses $ M_{k_i}$. Using the results in   eq.~(\ref{MkiSG}),  it follows that 
\begin{equation}
\label{SumaMi}  
M_{K(n,0)} =  \sum_{i=-\frac{n}{2}}^{\frac{n}{2}}  M_{k_i}  =\frac{8n^2}{n^2 -1 }  \cot  \frac{\pi}{2 n }.
\end{equation} 
To estimate  $\Delta V_{K(n,\alpha)}$, the integration domain $ \phi  \in  \{- n \pi ,  n \pi   \}$  is partitioned into $n$ subintervals of the form $ \phi  \in  \{ v_{i},  v_{i+1}   \}$. Within each subinterval, the smooth-modulation approximation is applied by taking
$1 + \cos (\frac{ \phi}{n} ) \approx 1 + \cos ( \frac{v_{i}  +v_{i+1}}{2n} ) $, hence    $\Delta V_{K(n,\alpha)}$ is readily  computed as follows:
  \begin{align}
\label{Vint}
 \Delta V_{K(n,\alpha)} \, &   \approx   \sqrt{8}  \sum_{i = 0  }^{  \frac{n}{2}  -1 }  \cos \left(  \frac{ (2i+1) \pi }{2n} \right)    \int_{ 2 i   \pi  }^{  2(i +1) \pi }   \left(  \sqrt{ \frac{ 1 + \alpha - \cos  \phi } {1 +  \frac{\alpha}{2} }} - \sqrt{1 - \cos \phi) }  \right) d\phi
     \\ \nonumber &=
    \sum_{i = 0  }^{  \frac{n}{2}  -1 }  \cos \left(  \frac{ (2i+1) \pi }{2n} \right)    \left(  \sqrt{32} \sqrt{ \frac{ \alpha  } {1 +  \frac{\alpha}{2} }}  \left( E \left[  (i+1)  \pi , -  \frac{ 2}{\alpha}\right] -   E \left[  i  \pi , -  \frac{ 2}{\alpha}\right]  \right) - 16 \right)
   \\ \nonumber &  =
  \left(    \sqrt{ \frac{ 32 \alpha  } {1 +  \frac{\alpha}{2}}} E \left[  -  \frac{ 2}{\alpha}\right] - 8   \right) \, \csc  \left( \frac{\pi}{2n}  \right).
\end{align}
The integrals involved in the computation are expressed in terms of the incomplete elliptic integral of the second kind, $E \left[  i\pi, -  \frac{2  }{\alpha}\right] =2 i \, E \left[ - \frac{2  }{\alpha}\right] $,
evaluated at an  argument  that simplifies the expression to a complete elliptic integral. This allows us to explicitly evaluate the sum, leading to the final result presented in  eq.~(\ref{Vint}). Figure~\ref{FigBen} $(a) $ compares this  analytical approximation  with the numerical integration of the second term in   eq.~(\ref{MKSM1}). The smooth-modulation approximation shows excellent agreement across the entire range of $\alpha$ values. 

 By adding $M_{K(n,0)}$ and  $\Delta V_{K(n,\alpha)}$, we obtain the total multi-kink mass:
 \begin{equation}    
 \label{MKSM2}
M_{K(n,\alpha)} \, = \left [ \frac{8n^2}{n^2 -1 } \cos  \left(  \frac{\pi}{2 n } \right)  + 
\sqrt{ \frac{ 32 \, \alpha }{1 +  \frac{\alpha}{2}  } }  E \left[ - \frac{2  }{\alpha}\right] -8 
 \right ]   \csc \left(  \frac{\pi}{2 n }    \right).
 \end{equation} 
Using the expansion   
  $\sqrt{ \frac{ 32 \, \alpha }{1 + \alpha } }  E \left[ - \frac{2  }{\alpha}\right] -8  \approx  \alpha  \left(  \ln  \frac{ 32}{ \alpha } -1  \right)  $  for  $ \alpha \ll 1$, the  small $\alpha$  limit of    eq.~(\ref{MKSM2}) yields    
 \begin{equation}    
 \label{MKSM3}
M_{K(n,\alpha)} \, \approx \left [ \frac{8 n^2 }{n^2 -1 }  \cos  \left(  \frac{\pi}{2 n } \right)  + \alpha 
\ln \left[  \frac{12  }{\alpha}\right]
 \right ]   \csc \left(  \frac{\pi}{2 n }    \right).
 \end{equation} 

  \subsubsection{Multi-kink  profile $\Phi_{n} (z)$  and size  $\sigma_{K_{(n,\alpha)}}$}
  Although the mass and charge soliton  are well defined for all positive values of $\alpha$, the emergence of a well-resolved multi-kink structure is restricted to the range $0 < \alpha < \alpha_1$, where $\alpha_1$ denotes the upper bound beyond which the internal sub-kinks can no longer be clearly distinguished. We now proceed to estimate the value of $\alpha_1$.

To begin, note that the separation between adjacent sub-kinks, $\xi_{i+1} - \xi_i$, can be inferred from the Bogomol'nyi equation, eq.~(\ref{Bogo}), under the assumption that each sub-kink is centered at the point where the field $\phi(z)$ reaches a local maximum of the potential, $i.e.$  $\phi(\xi_{i})=\Omega_{i}$. Accordingly \cite{Manton2019,Lozano2021},
 
  \begin{equation}
\label{IntEqmX2}
 \xi_{i+1}-\xi_{i} =  \int_{\Omega_{i}}^{\Omega_{i+1}} \frac{ d  \phi }{ \sqrt{2  \left[   V_{n,\alpha}(\phi)   \right] }}. 
  \end{equation}
Here   $\phi(\xi_{i} ) =  \Omega_{i}$ and   $\phi(\xi_{i+1} ) =  \Omega_{i+1}$ are contiguous maxima of the 
$ V_{n,\alpha}(\phi) $ potential. 
Recalling that in each region $\phi \in \{\Omega_{i-1}, \Omega_i\}$ the term $1 + \cos\left( \frac{\phi}{n} \right)$  can be approximated as $1 + \cos ( \frac{ v_i}{n})= 2m_{i}^2 $, hence a direct integration yields:
 
   \begin{equation}
\label{XeqSM1}
   \xi_{i+1} - \xi_{i}  \approx
  \frac{1}{2m_{i}}    \int_{\Omega_{i}}^{\Omega_{i+1}} \sqrt{ \frac{1 + \frac{\alpha}{2}  }{  \frac{ \alpha }{2} +  \sin^2 \frac{\phi}{2}   \ } }     d  \phi
    =  \frac{1}{m_{i}}  \sqrt{ \frac{2 (1+  \frac{\alpha}{2})}{ \alpha} } \,   \left(  F \left[  \frac{\Omega_{i+1}}{2}, -  \frac{2  }{\alpha}\right]   -  F \left[  \frac{\Omega_{i}}{2}, -  \frac{ 2}{\alpha}\right]   \right),
 \end{equation} 
where $ F \left[ \phi , k \right] $ is the incomplete  elliptic integral function of the first kind.   The previous  expression can be reduced  approximating  $\omega_{i} \approx (2i -1) \pi $, and using the property $F \left[  \frac{2i+1}{2}\pi,  -  \frac{2  }{\alpha} \right] =  (2i+1)K \left[ -  \frac{2  }{\alpha} \right]$, with $K \left[ -  \frac{2  }{\alpha} \right]$ being the complete elliptic integral function of the first kind,  then  eq.~(\ref{XeqSM1}) becomes  
 \begin{equation}
\label{XeqSM2}
   \xi_{i+1} - \xi_{i}  \approx
    \frac{2}{m_{i}}  \sqrt{ \frac{2 (1+   \frac{\alpha}{2} )}{ \alpha} } \,   K \left[ -  \frac{2  }{\alpha}   \right]   \approx   \frac{1}{m_{i}}   \log \left[ \frac{32}{\alpha} \right].
    \end{equation} 
Where the last result applies in the $\alpha  \ll 1$ limit. We highlight that the  $\log \left[ \frac{const}{\alpha} \right]$  dependence for the sub-kinks separation and $\alpha\, \log \left[ \frac{const}{\alpha} \right]$ for the 
multi-kink mass are generic results that can be deduced using the thin-wall approximation 
for scalar field models in which the vacuum degeneracy is partially broken by an small $\alpha$ parameter (see, $e.g.$ \cite{Christ1975,Brito2014,Demirkaya2017,Lozano2021,Dorey2023}).

The last result in   eq.~(\ref{XeqSM2}) allows us to estimate the condition required for a well-defined multi-kink structure, in which the internal kinks do not significantly overlap. From the analytical expression for $\phi_{k_i}(z)$ in   eq.~(\ref{phiAn}), the width $\sigma_{k_i}$ of the energy density associated with the $k_i$ kink can be estimated as
 $\sigma_{k_i} =\frac{ 8}{ m_{i-1} +m_i}$. 
To prevent overlap between adjacent kinks, we require that the separation between sub-kinks exceeds their width, i.e.,  $  \xi_{i+1} - \xi_{i}   > \sigma_{k_i}$,  leading to the following bound:
\begin{equation}
\label{alpha1}
 \alpha <   \alpha_1 =   32 \, \exp{\left \{- 8 \frac{ m_i}{ m_i+m_{i+1}} \right\} } \approx  0.59 \, .
 \end{equation} 
Notice that the previous  condition is  independent of the value of $n$. 
 
The previous results  enable us to construct an approximate expression for the $\Phi_{K_n}(z)$ profile.
We assume an ansatz in which the multi-kink is written as a superposition of the sub-kink solutions $\phi_{k_i}(z)$ given in  eq.~(\ref{XeqSM1}) separated by distances given in 
   eq.~(\ref{XeqSM1}). Hence  $\Phi_{K_n}(z)$ takes the following form  
  \begin{equation} 
\label{PhiK}
\Phi_{K_n}(z)\, = \ \sum_{i=1}^{\frac{n}{2}} \left(   \phi_{k_i} (z- X_i)  +   \phi_{k_{-i}} (z+ X_i) \right),   
 \end{equation} 
where  the internal kinks are centered at the  positions

 \begin{equation} 
\label{Xi}
X_i =   \sum_{j=1}^{i}  (1-  \frac{1}{2}   \delta_{j,1} )  \, (\xi_{j} - \xi_{j-1}) ,
 \end{equation} 
with $\xi_{0}  \equiv \xi_{-1}    $. The multi-kink   $\Phi_{K_n}(z)$ is centered at $z=0$.

\begin{figure}
 \begin{center}
\includegraphics[width=14.0cm]{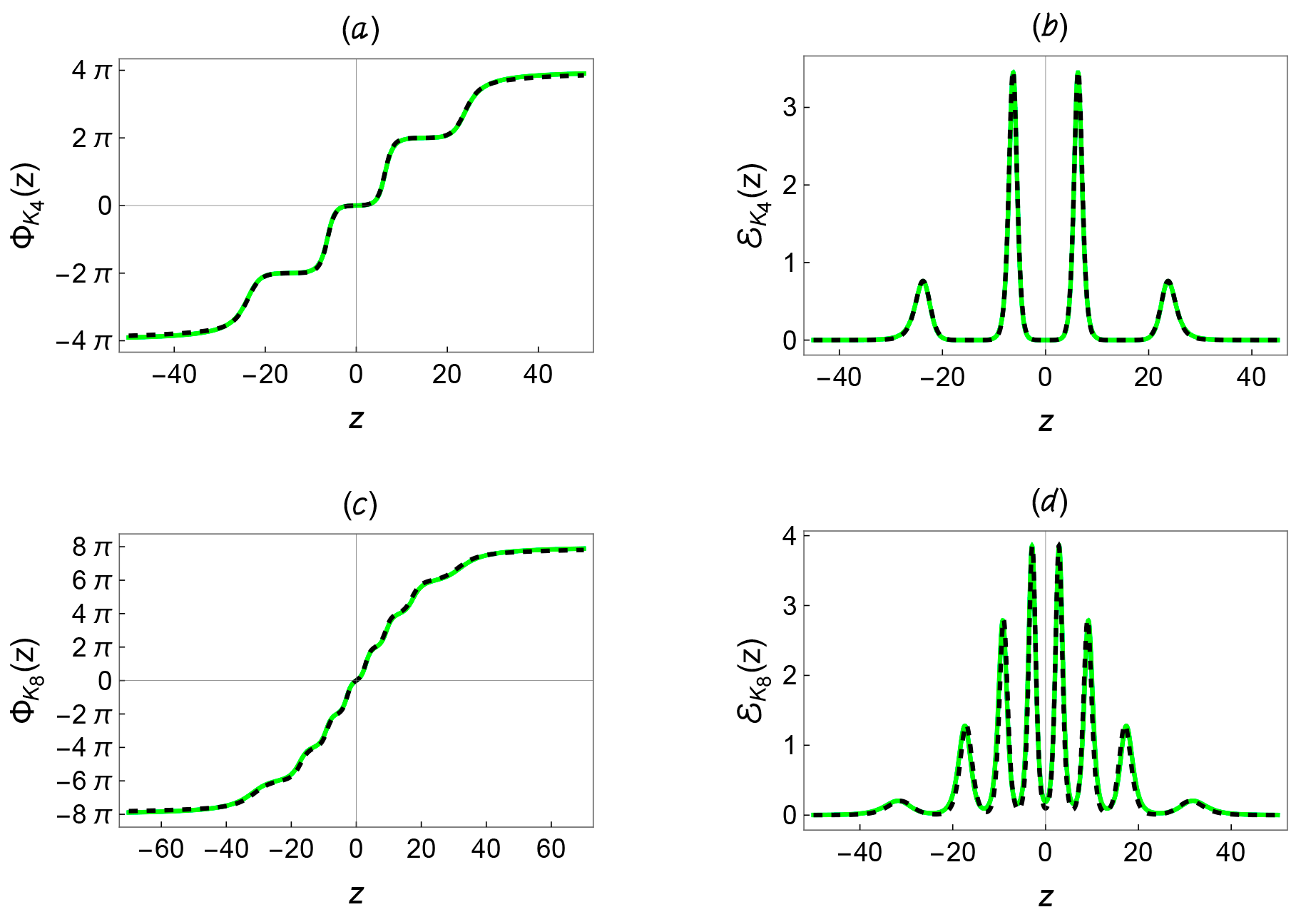}
\caption{(a) and (c) multi-kinks  $\Phi_{K_n}(z)$  obtained from the numerical solution  of  eq.~(\ref{Bogo}) for the potential $V_{n,\alpha} (\phi) $ in  eq.~(\ref{Pot2SG})    (continuous-green lines) and   the analytical  ansatz  eq.~(\ref{PhiK}) (dashed-black lines). (b) and (d) the corresponding   energy density  distributions. The parameters values are:  $n=4$, $\alpha=0.0001$ for the (a) and (b) figures and    $n=8$,  $\alpha=0.1$  in the  (c) and (d) cases.
  }
\label{FigPhin}
\end{center}
\end{figure}

Figure~\ref{FigPhin} compares the multi-kink profile obtained from the numerical solution of the exact Bogomol'nyi equation,  eq.~(\ref{Bogo}),  with the approximated solution obtained  from eq.~(\ref{PhiK}).
Importantly, the masses  $m_{\pm\frac{n}{2}}$  do not vanish any more, hence,  the expression for field $ \phi_{k_i} (z)$ in the left side of eq.~(\ref{phiAn}) applies for all values of the indices $i \le \frac{n}{2}$.
 The comparison shows excellent agreement between the analytical approximation and the numerical results, with the accuracy improving as $\alpha$ decreases and $n$ increases. The corresponding energy density plot reveals $n$ well-localized energy peaks, each associated with one of the $n$ sub-kinks, confirming the internal structure of the multi-kink configuration. 

  As established above, $K_n$ is composed of $n$ sub-kinks, and can thus be schematically represented as $K_n \sim \{k_{-{\frac{n}{2}} },k_{-{\frac{n}{2}} +1}, ........ ,k_{\frac{n}{2} }  \}$. 

It is reasonable to assume that the size of an individual sub-kink $k_i$ is characterized by the width of its energy density distribution, given approximately by $  \sigma_{k_i}  \approx \frac{8}{m_i+m_{i+1}}$. However, the total size of the multi-kink configuration must account for the full extent of its $n$ internal sub-kinks, as well as half the width of the two outermost sub-kinks, $k_{\pm \frac{n}{2}}$. Accordingly, we define the total  multi-kink size $\sigma_{K_{(n,\alpha)}}$  as

\begin{equation} 
\label{sigmaK}
 \sigma_{K_{(n,\alpha)}}  =   2 \left( X_ \frac{n}{2} + \frac{4}{m_{\frac{n}{2}-1}+ m_{\frac{n}{2}}} \right), 
 \end{equation} 
 where according to   eq.~(\ref{Xi})  $X_ \frac{n}{2}$ represents the position of the $k_{ \frac{ n}{2} }$ sub-kink.  
 Figure~\ref{FigBen} $(b)$  displays the multi-kink size as a function  of the  breaking degeneracy parameter $\alpha$  for various values of $n$. The result is compared with the approximated expression 
 
\begin{equation} 
 \label{sigmaK2}
\sigma_{K_{(n,\alpha)}}  \approx   \frac{2n}{\pi} \,\left[ \left( \ln \left( \frac{2n}{\pi} \right) +\frac{1}{2} \right)  \, \ln \left( \frac{32}{\alpha} \right) +4\right],
 \end{equation} 
   that is worked out considering  $\alpha \ll 1$,  $ n \gg 1 $ and using  the approximated result 
 $ \sum_{j=1}^{\frac{n}{2}-1} \frac{1}{\cos\left( j \pi/n\right)} \approx  \frac{n}{\pi} \left[ \ln\left( \frac{2n}{\pi} \right) + \frac{1}{2} \right] $ 
 in    eqs.~(\ref{XeqSM2},\ref{Xi}). Therefore we obtain  that the multi-kink size scales as 
  $ \sigma_{K_{(n,\alpha)}}  \sim   \frac{2n}{\pi} \, \ln(n)  \, \ln(\frac{1}{\alpha})$  as $n$ grows and   $\alpha$ decreases, reflecting the enhanced separation between sub-kinks in the weak-coupling regime.

\subsection{  Sub-kinks and multi-kink  spectra }\label{spect}
Considering a linear perturbation of the form  $\psi(z)=  \varphi (z)+ \eta(z) e^{i \omega t}$, around a soliton solution $ \varphi(z) $, the fluctuation spectrum is obtained from the eigenvalue equation

\begin{equation} 
\label{spectra}
\left[ -\frac{d^2}{dz^2} +U_{eff}(z) \right] \eta(z)  =\omega^2 \eta(z), 
 \end{equation} 
where the effective potential for the linear excitations is given by     $U_{eff}(z) = V^{" }_{n,\alpha} ( \varphi (z))$. 
Equation (\ref{spectra}) 
 was solved numerically for two cases:  (i) the $\alpha=0$ kinks $k_i$, and (ii) the multi-kink  $K_n$. In both cases, the resulting spectra contain only the translational zero mode and the continuum states, with no evidence of discrete internal (shape) modes.

\begin{figure}
 \begin{center}
\includegraphics[width=16.0cm]{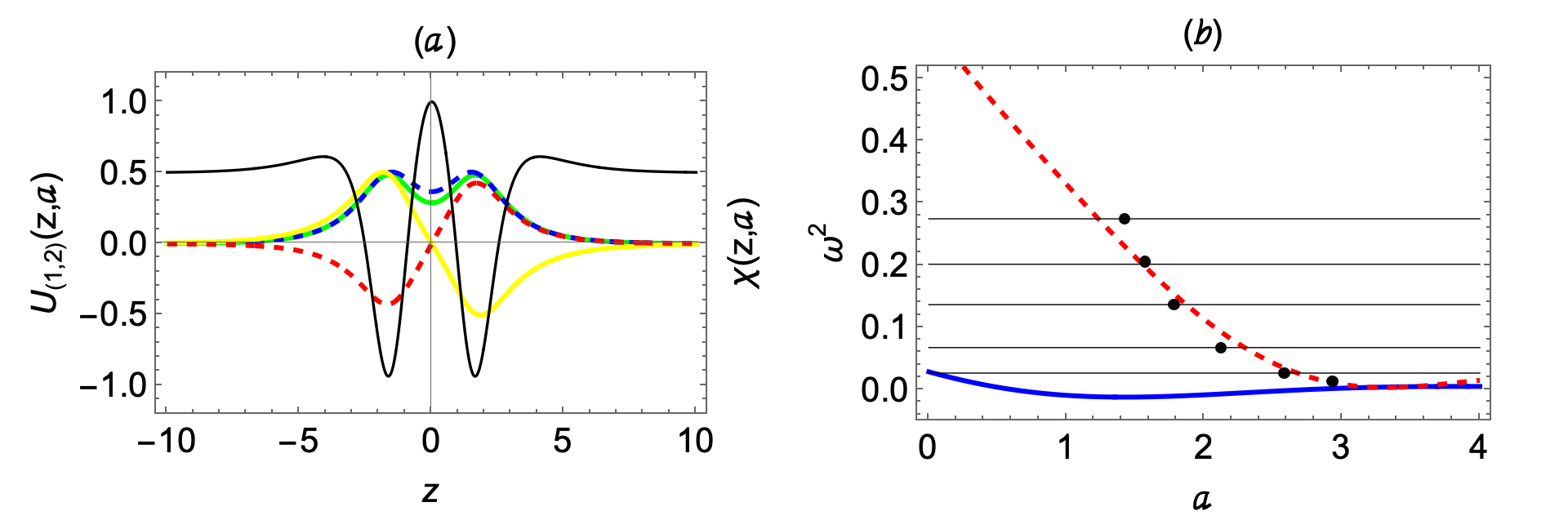}
\caption{ Fluctuation spectra around the $k_1-k_2$ sub-kink's pair of  $K_4$. $(a)$   
   $V^{" }_{n,\alpha} (  \varphi_{(1,2)}(z,a) )$ effective potential (black line) and the $ \eta_{S,A} (a,z)$ discrete modes obtained from the solutions of Equation~(\ref{spectra})  (continuous lines) and Equation~(\ref{eigenfSA})  (dashed lines) for $a=1.6$.   $(b)$  Eigenvalues for the symmetric quasi-zero mode  (blue solid line) and  anti-symmetric (red dashed line) bound states  as a function of the  sub-kink's separation $a$. The dots values are obtained from the equilibrium solutions $2a=\delta \xi_{\{1-2\}} $  eq.~(\ref{XeqG}),  and   $\omega$ eq.~(\ref{w1p1}) for   $\alpha $ values in the range $\alpha \in \{ 0.05-0.5\}$.  } 
\label{FigSm1}
\end{center} 
\end{figure}

Although no shape modes are found for the stationary $K_n$ multi-kink configuration, we expect the existence of internal vibrational modes associated with the oscillations of its constituent sub-kinks. To investigate this possibility, we take as the initial configuration an ansatz composed of two contiguous sub-kinks separated by a distance $2a$, given by

\begin{equation} 
\label{anza1}
   \varphi_{(i,i+1)}(z,a)=  \phi_{k_{i}}(z+a) + \phi_{k_{i+1}}(z-a) -v_i . 
 \end{equation} 
 Equation~(\ref{spectra}) was then solved using the effective potential    $U_{(i,i+1)}(z,a)=  V^{" }_{n,\alpha} (  \varphi_{(i,i+1)}(z,a) )$. 
 The results for  $n=4$ and $i=1$ are shown in figure~\ref{FigSm1}, where one observes that the effective potential exhibits a double-well structure, giving rise to two discrete modes. The first corresponds to a quasi-zero mode, whose eigenfunction
$ \eta_S(a,z)$  is approximately the symmetric combination of the zero modes of $k_i$ and $k_{i+1}$:

\begin{equation} 
\label{eigenfSA}
  \eta_{S,A} (a,z) =  \frac{1}{4} \left( \frac{d   \phi_{k_{i}}(z+a)} {dz }  \pm  \frac{d   \phi_{k_{i+1}}(z-a)} {dz } \right), 
 \end{equation} 
in contrast, the antisymmetric combination $ \eta_A(a,z)$ represents a genuine discrete mode trapped between the sub-kinks. This interpretation is  supported by the results discussed in section~\ref{forceip1}, where we determine the interaction force $F_{i,i+1}$, eq.~(\ref{Force}), acting between neighboring sub-kinks. The predicted values for the equilibrium separation and oscillation frequency,   eqs.~($\ref{XeqG}$, $\ref{w1p1}$),  are indicated by the circles in figure~\ref{FigSm1} (b), showing good agreement with the numerically obtained antisymmetric discrete mode in the small  $\alpha \le 0.5$ regime.

\begin{figure}
 \begin{center}
\includegraphics[width=16.0cm]{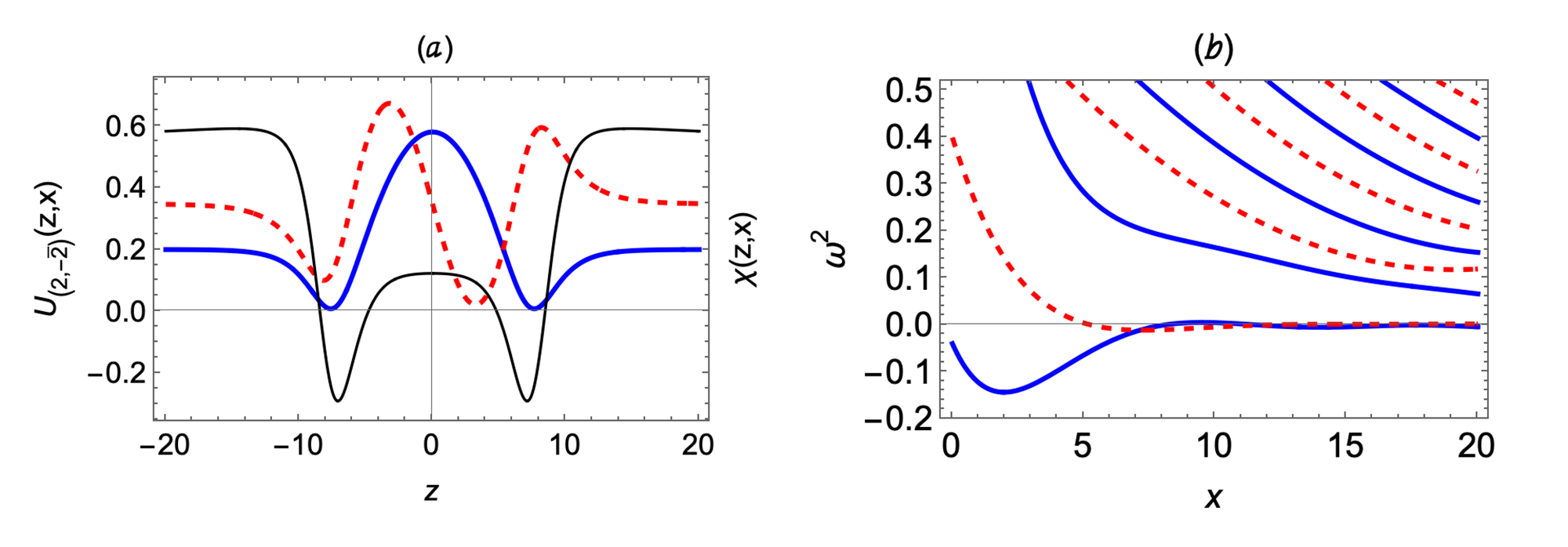}
\caption{ Fluctuation spectra around the  $k_{2}- {\bar k}_{-2 }$ sub-kink's pair  in a   ${K}_4-\,{\bar K}_4 $ collision.  $(a)$ $U_{( 2,-{\bar 2})}  (z,x)$  effective potential (black line) and the first two  discrete modes obtained from the solutions of Equation~(\ref{spectra}) for $x=7.5$.   $(b)$  Even  (blue solid line) and  odd  (red dashed line) bound states eigenvalues  as a function of the  sub-kink's separation $x$. }

\label{FigSm2}
\end{center}
\end{figure}

  In a collision  between the  multi-kinks ${K}_n \equiv \{k_{-{\frac{n}{2}} },, .... ,k_{\frac{n}{2} }  \} $ and  $  \,{\bar K}_n \equiv \{\bar {k}_{-{\frac{n}{2}} },, .... ,{\bar k}_{\frac{n}{2} }  \}$,  the dynamics is mainly governed by the interaction between the nearest contiguous sub-kinks, namely $k_{\frac{n}{2} } $ and $\bar {k}_{-{\frac{n}{2}} }$. In this case, the corresponding fluctuation spectrum is obtained by introducing the ansatz   
  \begin{equation}   
  \label{anza2}
\varphi_{( {\frac{n}{2}},-{\frac{\bar n}{2}})} (z,x) = \phi_{k_{\frac{n}{2}}}(z+x) + \phi_{\bar {k}_{-{\frac{n}{2}} }} (z-x)-v_{\frac{n}{2}} , 
 \end{equation} 
 and solving eq.~(\ref{spectra}) again, now for the effective potential     $U_{( {\frac{n}{2}},-{\frac{\bar n}{2}})}  (z,x) = V^{" }_{n,\alpha} ( \varphi_{( {\frac{n}{2}},-{\frac{\bar n}{2}})}  )$.  For this configuration, figure~\ref{FigSm2} shows that the effective potential develops a broad central well, which supports a larger number of internal modes. The number of such modes increases as the kink-antikink pair separates.

 The present analysis indicates that the standard procedure for deriving  kink collective coordinate equations of motion \cite{Sugiyama1979,Takyi2016,Manton21A,Manton21B}  is impractical here, since no analytical expression exists for the $\Phi _{K_ N}(z)$ multi-kink or their numerous vibrational shape modes, and those expected in the $K_n-{\bar K}_n$ collision. Instead, we adopt an alternative formulation in which the collective coordinates are described by a set of $n$ nonlinear coupled oscillators. The model is constructed once the inter-sub-kink forces and damping terms due to radiative losses are determined, providing a consistent framework for describing the multi-kink dynamics, as detailed in the following section.
 
  \section{Collective coordinate model }\label{SECCCM}

We construct a collective coordinate model to describe the dynamics of the multi-kink. Considering  that  the static ansatz $\Phi_{K_n}(z)$ in  eq.~(\ref{PhiK}), obtained by superposing $n$ sub-kinks, provides an accurate approximation to the exact solution, the  position $X_i$ of each  sub-kink ${k_i}$ is promoted to a dynamical variable $x_i(t)$. These coordinates evolve according to classical equations of motion under the forces exerted by the adjacent sub-kinks $k_{i-1}$ and $k_{i+1}$. Approximate expressions for these forces, as well as for the interaction between a $K_n$ multi-kink and its anti-multi-kink, are derived below.   
    
  \subsection{Forces between   $k_{i}- k_{i+1}$  sub-kinks}\label{forceip1}
  
Previously, Manton \cite{Manton2019} (see also \cite{Dorey2011}) developed an approach to analyze kink-antikink interactions characterized by long-range power-law tails, rather than the conventional exponential decay. In what follows, we extend this framework to the multi-kink case formulating a dynamical Bogomol'nyi relation. This generalization enables us to derive the effective forces $F_{{i,i+1}}$ acting between contiguous sub-kinks $\phi_{k_i}$ and $\phi_{k_{i+1}}$ that compose a multi-kink configuration $K_n$. These forces provide the foundation for the collective coordinate model developed in the next section.
  
   Let us assume that the  multi-kink configuration  $ \phi(z) $ satisfies the exact static field equation  eqs.~(\ref{eqphi1},\ref{Bogo}), except in the region  $\xi_{i}  \le  z \le \eta_{i+1}$,  where the field  $ \varphi(z,t) $ departs from the exact  $ \phi(z) $. Outside this region, the exact field  configurations  $ \phi(z) $ remains intact. The modified field  becomes time dependent and is  centered at $z=x_{i+1}(t) \equiv x(t) $, thus is  written as $ \varphi(z-  x(t) ) $, where   $x(t)$ represents the position of the sub-kink $k_{i+1}$. At equilibrium $ \varphi(z-  x(t) ) $  is expected to relax to the exact configuration  $ \phi(z) $ and   $x(t)   \to \xi_{i+1} $. Substituting $ \varphi(z- x(t) ) $, into the time dependent field equation  eq.~($\ref{eqphi}$) gives
\begin{equation}
\label{Eqccol}
    \gamma^{-2}   \frac{ d^2 \varphi}{d^2  y }  -     \frac{ d V_{n,\alpha} }{d  \varphi }
     + \ddot  x   \frac{ d \varphi}{d  y } =0  ,  
 \end{equation}
where $ y =z-  x(t) $ is the argument of $\varphi $  and   $ \gamma=1 /
    \sqrt{1-  \dot  x^2}$ is the Lorentz contraction factor.  Eq.~($\ref{Eqccol}$) can be considered a  equation for the field $\varphi (y) $  that evolves adiabatically on time trough its dependence  on $x(t)$.   In what follows we shall assume that  the acceleration  $\ddot  x $,  and the velocity   $\dot  x $ are   small, so only leading terms in these quantities will be retained. Multiplying    eq.~($\ref{Eqccol}$) by the term  $d \varphi / dy $, and utilizing  the relation $d \varphi/ dy = d W_{n,\alpha} / d\varphi$ leads to  the following expression 
  
 \begin{equation}
  \label{Bogow}
    \frac{ d }{d y }   \left(  \frac{ 1}{2 }  \left(   \frac{ \partial \varphi}{\partial  y }  \right)^2   - V_{n,\alpha}  +  \ddot  x  \,  W_{n,\alpha}\right) =0 , 
 \end{equation}  
  that is easily integrated from an arbitrary value of $z$ to  $z=\eta_{i+1} $, where $ \varphi(\eta_{i+1}) = v_{i+1}$, to give the following expression 
  \begin{equation}
\label{Bogo2}
   \frac{ 1}{2}  \left( \frac{ d  \varphi}{dy} \right)^2 =  \left[ V_{n,\alpha}(\varphi) -  \ddot  x  \left( W_{n,\alpha} (\varphi) - W_{n,\alpha} ( v_{i+1})\right) \right],   
 \end{equation}
 where we used the fact that $\varphi$  satisfy the  relation    
 $  \left[    \frac{ 1}{2}  \left( \frac{ d  \varphi}{dy} \right)^2 -  V_{n,\alpha}(\varphi)  \right]_{z =  \eta_{i+1}}    \!\!\!\!\!\!=0$  eq.~($\ref{Bogo}$), at the boundary. 
Comparing with  eq.~(\ref{Bogo}), we observe that  eq.~(\ref{Bogo2}) can be considered an adiabatic time dependent  Bogomol'nyi  relation, where the potential $V_{n,\alpha}(\varphi) $ is replaced by an effective potential  $V_{eff} =   V_{n,\alpha}(\varphi) -  \ddot  x \left( W_{n,\alpha} (\varphi) - W_{n,\alpha} ( v_{i+1})\right) $. 

For a given trajectory $x(t)$, the solution of   eq.~($\ref{Bogo2}$)  determines the adiabatic evolution of the field $\varphi$. Additionally, an integration of this equation throws an  expression for the evolution of the  coordinate $x(t)$. 
The  limits are selected from $z= \xi_{i}$    where  $\varphi=\Omega_{i}$, to $z=x(t)$   with $\varphi (x(t)) \to \Omega_{i+1}$, to obtain  
 \begin{equation}
\label{IntEqmX1}
 \int_{\Omega_{i}}^{ \Omega_{i+1}} \frac{ d  \varphi }{ \sqrt{2  \left[   V_{n,\alpha}(\varphi) +  {\ddot  x} \left( W_{n,\alpha}(v_{i+1}) -W_{n,\alpha}(\varphi) \right)  \right] }}   =    x(t)-\xi_{i}. 
  \end{equation}
This equation represents an implicit relation between the kink acceleration  ${\ddot  x}$ and $ x(t)$, that is the effective equation of motion for the coordinate $x(t)$. In particular, equilibrium is reached when $\ddot  x \to 0$ and $x(t) \to  \xi_{i+1} $,  reducing to the correct expression for the separation between the contiguous ${k_{i}}$  and ${k_{i+1}}$  kinks given in  eq.~($\ref{IntEqmX2}$).

 Approximated analytical expressions for the equation of motion for  $x(t)$ and the equilibrium separation  $\xi_{i+1}-\xi_{i}$  can be obtained from  eq.~($\ref{IntEqmX1}$). The leading contribution to the integral  in   eq.~($\ref{IntEqmX1}$) arises from  values of $\varphi$ in  the region close to the potential minimum $ \varphi \sim v_i$. Hence  we  expand the potential $V_{n,\alpha} (\varphi) = V_{n,0} (\varphi) + \alpha \Delta _{n,\alpha} (\varphi)  \approx  \frac{1}{2} m_{i}^2 (\phi - v_{i})^2  + \alpha \Delta _{n,\alpha}(v_{i})   $, where $m_{i}^2= (d^2V_{n,\alpha} / d^2 \phi  )_{\phi =v_{i}}$ is the mass of the  perturbative scalar oscillations around  $v_{i}$,  and  we verify that to leading order  $ W_{n,\alpha}(\varphi) \approx W_{n,\alpha}(v_{i})$.  The resulting expression is readily integrated,   and identifying  the $k_i$ kink  mass $ M_{k_{i+1}}=W_{n,\alpha}(v_{i+1}) -W_{n,\alpha}(v_{i})  $,   the relation between   $\ddot  x$ and  $x$ takes the form of a Newton equation of motion, 
   \begin{equation}
\label{Force}
M_{k_{i+1}}  \, \ddot  x_{i+1}  =  F_{\{i,i+1\}} \left(\xi_{i},  x_{i+1} \right) = 2 m_{i}^2 \left[ ( \Omega_{i+1} - v_{i} )  ( v_{i} - \Omega_{i}  )\right] e^{- m_{i} 
( x_{i+1} - \xi_{i} )}  -  \alpha \Delta _{n,\alpha}( v_{i}).
 \end{equation} 
The first contribution corresponds to the repulsive force that typically acts between kinks, while an additional attractive term arises from the negative pressure exerted by the false-vacuum region trapped between the kink's pair. At equilibrium, $x(t) \to \xi_{i+1}$, and the condition of vanishing total force determines the equilibrium separation as
 \begin{equation}
\label{XeqG}
\delta \xi_{ \{i,i+1\}} =  \xi_{i+1} - \xi_{i}= \frac{1}{m_{i}} \,  \ln  \left[   \frac{2 \left(  \Omega_{i+1}- v_{i} \right) \left( v_{i} - \Omega_{i} \right)}{ \alpha  }    \right],
 \end{equation} 
where,  to leading order in $\alpha$, we take into account that    $\Delta _{n,\alpha}(v_{i}) \approx  m_{i}^2$.  
The potential energy associated to $F_{\{i,i+1\}}$ is given as  
 \begin{equation}
 \label{Upot1}
 U_{\{i,i+1\} } \left( x_{i+1} -x_{i}  \right) = \alpha  m_i \left[ e^{ - m_{i}  \left[ y -  \delta \xi_{ \{i,i+1\}} \right] } -1 + m_i \left(  y -  \delta \xi_{\{i,i+1\}}    \right) \right]_{y=x_{i+1} -x_{i}}.
 \end{equation}
For small oscillations of the sub-kinks around their equilibrium separations, these results allow us to determine the corresponding oscillation frequency as
\begin{equation}
\label{w1p1}
 \omega^2_{i,i+1}  = \frac{ \kappa_{i,i+1}  }{ M_{k_i} }  \,,  \qquad \kappa_{i,i+1} =\alpha \, m_i^3 \,.
 \end{equation} 
 The previous results also apply for negative values of the indices, with the replacement $F_{-(i+1),-i}(a) \to - F_{i,i+1}(a)$.
 This extension includes the case of $F_{-1,1}(a)$, taking into account that $v_0=0$,  $\Omega_{-1} = -\Omega_{1}$ leading to  $\kappa_{-1,1} \equiv  \kappa_0 =\alpha \, m_0^3$. 
   
 The predicted values  for $\delta \xi_{ \{1,2\}}$ and $\omega_{1,2}$ in  eqs.~(\ref{XeqG}, \ref{w1p1}) are compared with the results obtained from  the fluctuation spectra obtained from the solution of    eq.~(\ref{spectra})  in figure \ref{FigSm1}, showing a good agreement in the 
 $\alpha \le 0.5$ regime. 

\subsection{Force between   $K_n$ and ${\bar K}_n$  multi-kinks}\label{forceKnAkn}

As previously mentioned, the $K_n-{\bar K}n$ interaction is dominated by the force between the nearest neighboring sub-kinks, $k_{\frac{n}{2}}$ and ${\bar k}_{-\frac{n}{2}}$. We now show that Bogomol'nyi-type relations, similar to eq.~(\ref{IntEqmX1}), together with information extracted from the full field simulations of the $K_n-{\bar K}_n$ collisions, see figure~\ref{Figbion}, allow us to derive an expression for the interaction force that incorporates both attractive and repulsive contributions, associated respectively with large and short separations between the colliding solitons.

For the $k_{\frac{n}{2}}-{\bar k}_{-\frac{n}{2}}$ field configuration, we consider the ansatz given in eq.~(\ref{anza2}). In this ansatz, the position of the left sub-kink is fixed at $x=x_0 <0$, while the antikink position is promoted to a dynamical variable $x(t)$. Consequently,  the effective field $\varphi(z-x(t))$ is expected to satisfy eqs.~(\ref{Eqccol}) and~(\ref{Bogow}). The field $\varphi$  
describes the evolution of the sub-kink within the region  $\{  v _{ \frac{n}{2} -1} - \pi  \le  \varphi  \le v _{ \frac{n}{2} } + \pi  \}$.

Let us first consider the large-separation regime. From figure~\ref{Figbion}, we observe that $\varphi(z-x(t))$ is localized around the absolute minimum $v_{\frac{n}{2}}=n\pi$ of the potential.  It is therefore convenient  to integrate eq.~(\ref{Bogow}) from an arbitrary value of $z$ to $z_r$, where $\varphi(z_r)=v_{\frac{n}{2}-1}$. This procedure yields an expression analogous to eq.~(\ref{Bogo2}), with the replacement $v_{i+1}\to v_{\frac{n}{2}-1}$. Performing an additional integration, we obtain the following expresion:

 \begin{equation}
\label{IntEqmX3}
 \int_{ (n-3) \pi }^{  (n+1) \pi } \frac{ d  \varphi }{ \sqrt{2  \left[   V_{n,\alpha}(\varphi) +  {\ddot  x} \left( W_{n,\alpha}( (n-2) \, \pi ) -W_{n,\alpha}(\varphi) \right)  \right] }}   =    x(t) + x_0. 
\end{equation}
The leading contribution to the integral arises from field values in the vicinity of $\varphi \sim n\pi$. This justifies the approximations  $V_{n,\alpha} (\varphi)  \approx  \frac{1}{2} m_{ \frac{n}{2}}^2  (\varphi - n \, \pi)^2 $, 
 and  $W_{n,\alpha} (\varphi)   \approx W_{n,\alpha } ( n \pi )   $. Under these assumptions, eq.~(\ref{IntEqmX3}) reduces to

\begin{equation}
\label{IntEqmX4}
 \int_{ (n-3) \pi }^{ (n+1) \pi } \frac{ d  \varphi }{  \sqrt{   m_{ \frac{n}{2}}^2   (\varphi - n \pi )^2    -   2\, {\ddot  x} \,  M_{ \frac{n}{2}} }}  =    x(t) + x_0,   
\end{equation}
where we have identified the sub-kink mass as $  M_{ \frac{n}{2}} =W_{n,\alpha } ( n \pi )  - W_{n,\alpha}( (n-2) \, \pi )  $. 
Since the acceleration is expected to be negative, $\ddot{x} = -|\ddot{x}|$, the integral is well defined and can be evaluated straightforwardly. This yields the following expression for the long-range interaction force $ M_{ \frac{n}{2}} \,   \ddot  x = -6   \pi^2   m_{ \frac{n}{2}}^2 \, \exp\left(-3 m_{ \frac{n}{2}}  \, (x + x_0 ) \right) $.


We now consider the regime in which the sub-kinks are in close proximity. Figure~\ref{Figbion} $(b)$ shows that the field amplitude near the central region ($z \sim 0$) is significantly reduced, fluctuating around the false vacuum $v_{\frac{n}{2}-1} = (n-2)\pi$, rather than around the true vacuum $v_{\frac{n}{2}} = n\pi$.
This observation motivates repeating the previous analysis, but now expanding around $v_{\frac{n}{2}-1}$. In practice, this corresponds to implementing the substitutions
$v_{ \frac{n}{2} -1}  \Longleftrightarrow v_{ \frac{n}{2} } $  and  $m_{ \frac{n}{2}}  \Longleftrightarrow m_{ \frac{n}{2}-1}$ within the integrals  appearing in eqs.~(\ref{IntEqmX3}) and~(\ref{IntEqmX4}). Proceeding in this way, we obtain

\begin{equation}
\label{IntEqmX5}
 \int_{ (n-3) \pi }^{ (n+1) \pi } \frac{ d  \varphi }{  \sqrt{   m_{ \frac{n}{2}-1 }^2   (\varphi - (n-2) \pi )^2    +   2\, {\ddot  x} \,  M_{ \frac{n}{2}} }}  =    x(t)  + x_0.  
\end{equation}
This expression is well defined for $\ddot{x}>0$, corresponding to a positive acceleration and therefore signaling the presence of a repulsive interaction. Explicitly, the repulsive force is given by $ M_{ \frac{n}{2}} \,   \ddot  x = 6   \pi^2   m_{ \frac{n}{2}-1}^2 \, \exp\left(-3 m_{ \frac{n}{2}-1}  \, (x+x_0) \right) $.

By combining both the large-separation (attractive) and short-range (repulsive) contributions, we  obtain the following:

 \begin{equation}   
\label{FKaK2}
F_{k_{\frac{n}{2}},\bar {k}_{-{\frac{n}{2}} }} (x)= 6 \,  \pi^2 \, \left ( m_{ \frac{n}{2}-1}^2  \,  e^{- 3 \, m_{\frac{n}{2}-1} \, (x+x_0) }-  m_{ \frac{n}{2}}^2  \,  e^{- 3 \, m_{\frac{n}{2}} \, (x+x_0) }  \right).
\end{equation}  
The value of $x_0$ can be determined by observing that, in the closest-collision regime, the $k_{\frac{n}{2}}$ and ${\bar k}{-\frac{n}{2}}$ sub-kinks oscillate around the origin. Therefore, imposing the equilibrium condition $x=x_{eq}=0$ fixes $x_0$ as

 \begin{equation}   
\label{x0}
x_0   =  \frac{ 1}{ 3 \,  \left( m_{\,  \frac{n}{2}-1} -m_{\frac{n}{2}}   \right)}  \, \log \left( \frac{m_{\frac{n}{2}-1 } }{m_{\frac{n}{2} }}  \right)^2  \, , 
\end{equation} 
 which leads to the final expression for the $k_{\frac{n}{2}}-{\bar k}_{-\frac{n}{2}}$  interaction force
 
 \begin{equation}   
\label{FKaK2}
 F_{k_{\frac{n}{2}},\bar {k}_{\!\!-{\frac{n}{2}} \!\!}}  (x) = C_{n,\alpha}\, \left (  e^{- 3 \, m_{\frac{n}{2} \!\! -1} \, x } \!\!\! -    e^{- 3 \, m_{\frac{n}{2}} \, x } \right), \quad  \quad C_{n,\alpha}   = 6 \, \pi^2 \, m^2_{\frac{n}{2}} \, \exp \left[  -3  \,   m_{\frac{n}{2}} x_0  \, \right]. 
\end{equation}
We recall that, according to eq.~(\ref{msc}), the  scalar meson mass associated with fluctuations around the global minima behaves as $m_{\frac{n}{2}} \sim \sqrt{\alpha}$, and is therefore significantly smaller than the masses associated with fluctuations around local minima, namely  $m_{\frac{n}{2}-1}$. Consequently, eq.~(\ref{FKaK2}) describes a Morse-type interaction, characterized by a long-range attractive component with characteristic scale $1/m_{\frac{n}{2}}$,  together with a short-range repulsive component with scale $1/m_{\frac{n}{2}-1}$. The corresponding interaction potential is given by

       \begin{equation}   
\label{Upot2}
U_{k_{\frac{n}{2}},\bar {k}_{-{\frac{n}{2}} }}  (x) =  \,  \frac{1}{3}  \, C_{n,\alpha}\, \left ( \frac{e^{- 3 \, m_{\frac{n}{2}-1} \, x }}{m_{\frac{n}{2}-1}} -    \frac{e^{- 3 \, m_{\frac{n}{2}} \, x }}{ m_{\frac{n}{2}}  }      \right) \, ,
\end{equation}  
while oscillations around the potential minimum have a frequency given by
 \begin{equation}
\label{wKAK} 
 \omega^2_{\frac{n}{2}, \frac{\bar {n}}{2} }  = \frac{ \kappa_{\frac{n}{2}, \frac{\bar {n}}{2} } } {M_{ \frac{n}{2}} }  \, ,         
 \quad    \quad  \varkappa_{\frac{n}{2}, \frac{\bar {n}}{2} }  = 3 \,  (m_{\frac{n}{2}-1} -m_{\frac{n}{2}}) \,  C_{n,\alpha}  \, .
  \end{equation} 
 

 An advantage of using the Bogomol'nyi-type relations, eqs.~(\ref{IntEqmX4}) and~(\ref{IntEqmX5}), is that they do not require explicit expressions for the soliton field configurations. The ansatz in eq.~(\ref{anza2}) was introduced only to define the relevant kink and antikink coordinates.

A more precise determination of the $K_n-{\bar K}_n$ interaction could, in principle, be obtained within the moduli space formalism~\cite{Manton21A,Manton21B}. However, this would require a more accurate characterization of the field configuration, taking into account both the short-distance behavior and the asymptotic regime of large separations between the colliding solitons, features that are not incorporated into eq.~(\ref{anza2}). For the purposes of the present work, we consider that the expression given in eq.~(\ref{FKaK2}) provides a sufficiently accurate description of the interaction. 
 
  
   \subsection{Radiation energy losses }\label{radsec}
  
   Previous investigations have mainly attributed radiation effects to nonlinear mechanisms. Examples include: (i) the decay of a single kink shape mode through the excitation of higher harmonics of a fundamental frequency lying below the scalar mass threshold~\cite{Manton1997,Navarro2023}; and (ii) nonlinear radiation mechanisms in oscillons~\cite{Gleiser1994,Rompineve2021}, including the emission of staccato radiation bursts when monotonically increasing frequencies cross the mass radiation threshold~\cite{Dorey2020}. The consequences of these mechanisms for the present study will be discussed in section~\ref{KnAKnsec3}. Here, however, we focus our attention on a linear radiation scheme, which becomes essential for explaining oscillon stability as well as the synchronization of its vibrational components.

 The  $K_n-{\bar K}_n$ collision spectra  displays an  ample range of characteristics frequencies 
  $\omega_l$. Although many of these frequencies  lie below the scalar mass thresholds $\omega_l  <  m_i$, there is a significant number of frequencies located above the $m_{\frac{n}{2}}$ threshold that contribute to the radiative energy losses.   This feature makes it possible to establish a first-order perturbative scheme in which back-reaction radiation effects are incorporated into an effective force that modulates the dynamical evolution of the collective coordinates ${x_i(t)}$.  
      
 Suppose we  consider the field  $\varphi(t,z)  = \phi_{k_i} (z- x_i(t)) + f_i(t,z)$, 
 where  $ f_i(t,z)$  denotes the radiation field  emitted by the $i$-th sub-kink and $ \phi_{k_i} (z)$ is the kink-solution  in the $\alpha=0$ limit. Substituting this ansatz into the field equation of motion eq.~(\ref{eqphi}),   and expanding to leading order in $f_i(t,z)$, we obtain  
  \begin{equation}
\label{eqrad1}
  \left(   \frac{\partial^2}{\partial t^2}-\frac{\partial^2}{\partial z^2}  - m_i^2\right)  f_i(t,z) = 
   \ddot  x_{i} \frac{\partial   \phi_{k_i} }{\partial  \xi},
\end{equation}
 where  we neglected terms of order $ {\dot  x}^2_i$ and $\alpha \frac{\partial \Delta_{n,\alpha} }{\partial \phi} $, and approximate 
  \begin{equation}
  \label{eqrad12}
\frac{\partial^2 V_{n,\alpha} [ \phi_{k_i} (\xi)] }{\partial \phi^2}  \approx m_i^2.
\end{equation}
  Henceforth we denote $v_i = \dot{x}_i$ and $a_i = \ddot{x}_i$. The solution can then be obtained using the one-dimensional Klein-Gordon Green's function, which yields~\cite{Campbell1983,Smyth1999} 
  \begin{equation}  
  \label{eqrad2}
 f_i(t,z) = \frac{1}{2} \int_{t_1} dt^\prime  \int dz^\prime \Theta (t -t^\prime -\vert z - z^\prime \vert) J_0 \left(m_i   \sqrt{(t -t^\prime)^2 - ( z - z^\prime )^2 } \right)
  a_{i} (t^\prime) \frac{\partial   \phi_{k_i} }{\partial  \xi^\prime},
\end{equation}  
where $\xi' = z' - x_i(t')$, $\Theta$ denotes the Heaviside step function, and $J_0$ is the Bessel function of the first kind. The last term in this equation is strongly localized around the sub-kink position and can therefore be approximated as $  \frac{\partial   \phi_{k_i} (z^\prime - x_{i} (t^\prime)) }{\partial  \xi\prime}  \approx  2 \pi  \delta_D (z^\prime - x_{i} (t^\prime))$, where $ \delta_D$ is the Dirac delta. Under this approximation,    eq.~(\ref{eqrad2}) simplifies to
  \begin{equation}  
  \label{eqrad3}
 f_i(t,z) = \pi \int_{t_1} dt^\prime \,  \Theta (t -t^\prime -\vert z -  x_{i} (t^\prime) \vert)  \, J_0 \left(m_i   \sqrt{(t -t^\prime)^2 - ( z -  x_{i} (t^\prime) )^2 } \right)  \, a_{i} (t^\prime).  
\end{equation}  

 The radiated energy per unit time  is obtained from the Poynting  term 
   \begin{equation}  
  \label{eqrad35}
S_i[t,z_R] =  \left[ \frac{\partial f_i}{\partial t}  \frac{\partial f_i}{\partial z} \right]_{z=z_R}, 
\end{equation}  
where $z_R $ is an observation point located at the radiation region  $z _R >  x_{i} $.  Differentiating the radiation field with respect to $t$ and $z$, and assuming that $  z \gg   x_{i} $,  one finds

 \begin{align}
\label{eqrad4} 
  \frac{\partial f_i  }{\partial t}  &=  \pi    a_i (t-z) - \pi \,    \int_{t_1}^{t-z}  dt^\prime \,  \,  \frac{m_i  (t-t^\prime) }{ \sqrt{ (t-t^\prime)^2 -z^2 }}  \, J_1 \left(m_i  \sqrt{ (t-t^\prime)^2 -z^2 }  \right)   \,  
   a_i (t^\prime)  \, ,  \\ \nonumber 
  \frac{\partial f_i  }{\partial z}   &=  -\pi    a_i (t-z)  + \pi \,     \int_{t_1}^{t-z}  dt^\prime \,  \,  \frac{m_i  z}{ \sqrt{ (t-t^\prime)^2 -z^2 }}  \, J_1 \left(m_i  \sqrt{ (t-t^\prime)^2 -z^2 }  \right)   \,  
   a_i (t^\prime)   \, .
\end{align}

As previously discussed, the acceleration spectra $a_i(\omega)$ are expected to contain a broad range of frequencies. In contrast, the field-radiation spectra  $f_i(\omega,z_R)$ should include only propagating modes with $\omega > m_i$. This implies that the Bessel functions appearing in eqs.~(\ref{eqrad3}) and~(\ref{eqrad4}) effectively act as frequency filters. This interpretation is fully consistent with the numerical results presented   in section~\ref{eqsec}, as shown in figures~\ref{Figpf34}.

For the internal sub-kinks involved in a  $K_n$-$\bar{K}_n$ collision,   figure~\ref{Figallacel} ($a$)  shows that shortly after the collision their acceleration spectra display a broad set of characteristic frequencies. Nonetheless, two sharply contrasting patterns emerge. For the internal sub-kinks, virtually all frequency modes lie below the corresponding mass threshold $m_i$, whereas the external sub-kink exhibits a significant fraction of modes above the radiation threshold. Consistent with this behavior, the  $ f_i (\omega)$ spectra in  figures~\ref{Figpf34}
shows a strong radiation suppression for the internal sub-kinks, while nearly all radiation originates from the external sub-kinks.

 A complete treatment of these effects would require solving a complicated system of coupled nonlinear integro- differential equations, which lies beyond the scope of the present work. However, the previous results motivate a simple self-consistent approximation that can be implemented by proposing the relations
\begin{equation}
 \label{eqrad45}
 \left [ \frac{\partial f_i (t) }{\partial t}   \right ]_{ap} \approx  \pi   \,  \sqrt{ \beta_i(t) }   \,  a_i (t) \, , \qquad   \left [ \frac{\partial f_i (t) }{\partial z}  \right ]_{ap} \approx   - \pi  \,   \sqrt{ \beta_i(t) }  \,    a_i (t)  \, .
\end{equation} 
Here $\beta_i(t)$  is  a slow varying  time-dependent function introduced to ensure that, on average, the total radiated power is incorporated correctly. The function  $\beta_i(t)$  is therefore determined by requiring that, over successive time intervals, the radiation energy losses computed from eqs.~(\ref{eqrad4}) and~(\ref{eqrad45}) remains mutually consistent 
\begin{equation}
\label{eqrad5}
\int_0^{\omega_{\text {max} } }   \left [ \frac{\partial f_i  }{\partial t} (\omega)
  \right ] \,  \left [ \frac{\partial f_i  }{\partial z} (\omega)
  \right ] \, d \omega  \, \sim \,  \int_0^{\omega_{\text {max} } }   \left [ \frac{\partial f_i  }{\partial t} (\omega)
  \right ]_{ap} \,  \left [ \frac{\partial f_i  }{\partial z} (\omega)
  \right ]_{ap}  \, d \omega  \, .
  \end{equation}
Under this approximation, the Poynting term in eqs.~(\ref{eqrad35}) reduces to the simple expression
   \begin{equation}  
  \label{eqrad55}
S_i[t] =  -   \pi^2 \, \beta_i (t)  \,  a_i(t)^2 \, . 
\end{equation}  
This result allows us to obtain an explicit expression for the back-reaction force $F_{\text{rad},i}$ by requiring that the work done by this force matches the radiated energy loss over the same time interval. This  results in  an Abraham-Lorentz type  expression  
\begin{equation}  
  \label{eqrad6}
 F_{rad,i} =  \,   \   \pi^2 \, \left( \beta_i   \frac{d a_i} {dt} \, +   a_i  \frac{d  \beta_i  } {dt} \,    \right) . 
\end{equation}   
The first term corresponds to the effective damping acting on the $i$-th sub-kink, while the second term yields a correction to its effective mass.
   
    \subsection{Equations of motion }\label{eqsec}
  
 Finally, collecting the previous results we can  establish the collective coordinate model that describes the $ K_n -{\bar  K}_n$ collisions. From now on, we consider the $n=4$ case,  its extension to arbitrary $n$ is straightforward. The equations of motion govern the evolution of the sub-kink components of   ${\bar K}_4 \sim  \{ {\bar k}_{-2}, {\bar k}_{-1}, {\bar k}_{1},  {\bar k}_{2} \}$,  while the corresponding components of $ K_4$ are obtained by imposing the $z \to -z$ symmetry.

In this model, only interactions between a sub-kink  and its nearest neighbors are taken into account. Consequently, the equations of motion  take the following form
 \begin{equation}  
  \label{ccm1}
 M_{k_i} {\ddot x}_i  \,  =  \,  {\cal F}_i (x_{i-1},x_{i},x_{i+1})\, ,  \qquad i=-2,-1,1,2,
\end{equation}   
 where $ {\cal F}_i $ is constructed using  eqs.~(\ref{Force},\ref{FKaK2}) and  leads to:  

  \begin{align}
\label{ccm2} 
&  { \cal F}_{-2} (x_{-2},x_{-1}) =   F_{\{K_2-\bar K_{2}\}}    (x_{-2}) +F_{\{1,2\}} (x_{-1} -x_{-2}) \, ,   \\ \nonumber 
&   { \cal F}_{-1} (x_{-2},x_{-1}, x_1) =   -F_{\{1,2\}} (x_{-1} -x_{-2}) -  F_{\{-1,1\}} (x_1 -x_{-1} )  \, ,   \\ \nonumber 
&    { \cal F}_1 (x_{-1}, x_1,x_2) =   F_{\{-1,1\}} (x_1 -x_{-1} ) -  F_{\{1,2\}} (x_2 -x_1) \, ,    \\ \nonumber 
&    { \cal F}_2 (x_1,x_2) =   F_{\{1,2\}} (x_2 -x_1) \,.
\end{align}
 It is straightforward to verify that the equations of motion,   eqs.~(\ref{ccm1},\ref{ccm2}), follow from the Euler-Lagrange equations applied to the Lagrangian function 
 $L \left[x_i , {\dot x}_i  \right] = T  -U \left[ x_i  \right] $, where the kinetic energy and  potential terms  are given as follows:
 \begin{align}
\label{Lagran} 
& T =  \sum_{i=-2 \, , i \neq 0}^{i=2} \, \frac{1}{2}   M_{k_i} {\dot x}_i^2  \, ,\\    \nonumber 
  U \left[ x_i  \right]=  
U_{K_n, \bar K_n}(x_{-2} ) + &  U_{\{1,2\} } \left( x_{-1} - x_{-2}  \right)  + U_{\{-1,1\} } \left( x_{1} - x_{-1}  \right) + U_{\{1,2\} } \left( x_{2} - x_{1}  \right) \, ,
 \end{align}
where the interaction  potential  $ U_{\{i,i+1\} } $ and $U_{K_n, \bar K_n}$  are defined in   eqs.~(\ref{Upot1},\ref{Upot2}). These results allow us to verify that, to a high degree of precision, the numerical solutions of the  equations  of motion  eq.~(\ref{ccm1}) are consistent with the conservation of the total energy $E_{CCM} =T + U$.

Back-radiation effects on the sub-kinks can be incorporated by adding the force term $F_{{\text{rad},i}}$ from eq.~(\ref{eqrad6}) to the equations of motion. However,  a direct substitution of terms involving the time derivative of the acceleration leads to the well-known issue of runaway solutions. To avoid this, we adopt the perturbative method of Landau and Lifshitz, in which the first-order solution of $ M_{k_i} {\ddot x}_i  \,  =  \,  {\cal F}_i$
 is used to evaluate the acceleration time derivative term appearing in  eq.~(\ref{eqrad6})  and thereby estimate $F_{\{ rad,i\}}$. Taking these considerations into account, the equations of motion take the following form:
  \begin{equation}  
  \label{ccm3}
{ \tilde M}_{k_i} {\ddot x}_i  \,  =  \,  {\cal F}_i (x_{i-1},x_{i},x_{i+1})\,   + \pi^2 \,\, \frac{  \beta_i(t) }{ M_{k_i}}    \sum_j  \,  (v_j -v_i) \,  \left [  \frac{ \partial {\cal F}_j (\xi) }{ \partial  \xi} \right]_{\xi= (x_j -x_i)}    \, ,
\end{equation}   
where  $ { \tilde M}_{k_i} = M_{k_i} -\pi^2  \frac{d  \beta_i  } {dt}. $  With the inclusion of radiative effects, the total particle energy $E_{CCM} =T + U$  is no longer conserved. However, one can verify that the energy loss $\Delta E_{\text{CCM}}$ over a given time interval is consistent with the radiated energy obtained from the time integral of the Poynting term
  \begin{equation}  
  \label{ConsEner}
 E_{CCM} (t_2) -E_{CCM} (t_1) \approx  E_{rad} (t_1,t_2) =  \sum_i \int_{t_1}^{t_2}  \, \beta_i (t) S_i[t,z_R] \,  dt \, . 
\end{equation}   

A consistent solution of eqs.~(\ref{ccm3}) requires knowledge of the function $\beta_i(t)$. To determine it, we employ a recursive approximation scheme. In the first step, we set $\beta_i \equiv 0$ and solve the equations of motion (\ref{ccm1}). Once the trajectories $x_i(t)$ and their time derivatives are obtained, all quantities entering eqs.~(\ref{eqrad3}), (\ref{eqrad35}), and (\ref{eqrad4}), as well as the required Fourier transforms, can be evaluated. Then, for a sequence of post-collision time intervals $t_{pc} \in \{t_{pc}-\delta t,, t_{pc}+\delta t \}$, values  of $\beta_i(t_{pc})$ are  determined from  eq.~(\ref{eqrad5}). Two or three iterations of  the process, solving   eqs.~(\ref{ccm3}),   leads  to convergent results, $e.g.$ for  $\alpha = 0.05$ and $v_i = -0.1$:

\begin{align}
\label{beti2} 
\beta_i(t) & \approx \frac{ { \tilde \beta_i} }{\pi} \left(  \arctan  \left[ -m_{eff} \left( t - \left(\frac{X_L}{\vert v_i \vert }+ t_c  \right) \right) \right]  +\frac{\pi}{2 }\right) \, , \,\,\,\,\,  m_{eff} \approx   0.0002 \,,  \,\,\,\,\, t_c \approx 6400 \, , \\ \nonumber 
{\tilde \beta}_{-2} & \approx 0.00045 \, , \qquad  {\tilde \beta}_{-1} \approx  0.003  \, ,  \qquad  {\tilde \beta}_{1}   \approx  0.003  \,,  \qquad  {\tilde \beta}_{2}   \approx  0.7.  
 \end{align}
  The extracted ${\tilde{\beta}}_i$ values confirm that radiation is strongly suppressed for internal sub-kinks and originates primarily from the external one. The time dependence of $\beta_i(t)$ captures the early energy loss during the first rebounds, followed by stabilization into a long-lived oscillon. Although obtained for a particular selection of  $\alpha $ and $v_i $,  the results in eqs.~(\ref{beti2}) vary only weakly within the ranges $\alpha \in [0.001, 0.5]$ and $v_i \in [-0.01, -0.5]$.
 \begin{figure}
 \begin{center}
\includegraphics[width=14.0cm]{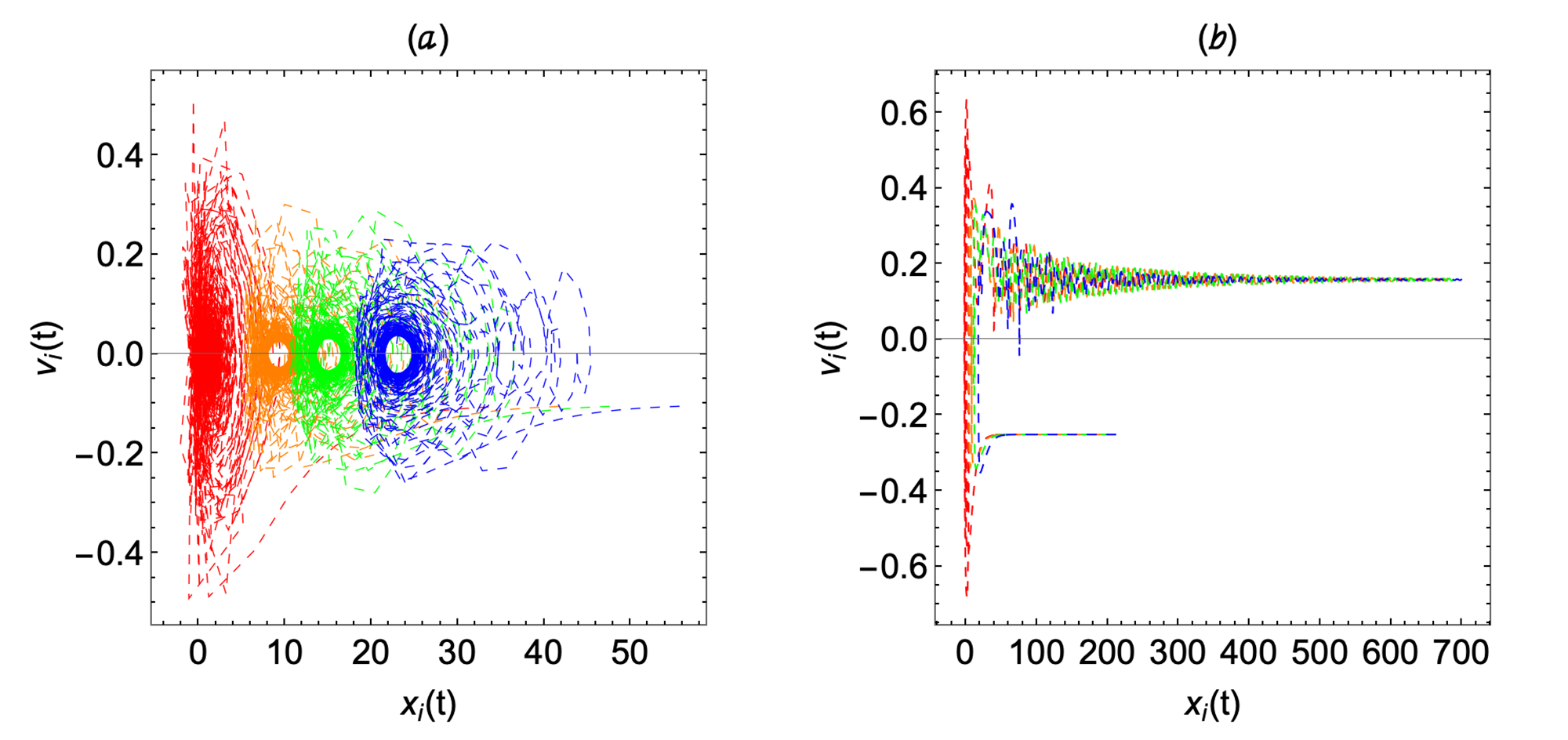}
\caption{ The  $\left(x_i,\dot{x}_i \right)$  phase  space portrait for $K_4-{\bar K}_4$ collision solutions
 of eqs.~(\ref{ccm3}).  The trajectories correspond to the:  ${\bar k}_{-2}$   (red line),   ${\bar k}_{-1}$  (orange line),   ${\bar k}_{1}$ (green  line),  and  ${\bar k}_{2}$  (blue line) sub-kinks. $(a)$ Oscillon evolution in the $t \in \{1500,100000 \}$ time interval,  for $v_{in}=-0.1$. $(b)$ Inelastic dispersion for $t \in \{0,15000 \}$,  when $v_{in}=-0.25$. 
Here $n=4$,  and  $\alpha=0.05$.
  }
\label{FigPhaseS}
\end{center}
\end{figure}
      
The $(x_i,\dot{x}_i)$ phase-space portrait for a solution of eqs.~(\ref{ccm3}) is shown in figure~\ref{FigPhaseS} ($a$). The evolution spans a long interval, $t \in [0,190,000]$, and is consistent with the formation of a long-lived oscillon: the oscillation amplitude decreases significantly during the first rebounds, and for $t > 10,000$ the $(x_i,\dot{x}_i)$ trajectories settle around slowly contracting elliptic limit cycles, as expected for a metastable configuration with weak radiative damping. A detailed comparison with the corresponding full field simulation will be presented in the next section. The energy balance also confirms the consistency of the collective coordinate model, with
$E_{CCM} (20000) -E_{CCM} (0) = -0.8905$ and $E_{rad} (0,20000) =  -0.8946$, 
showing excellent agreement with eq.~(\ref{ConsEner}).

\begin{figure}
 \begin{center}
\includegraphics[width=16.0cm]{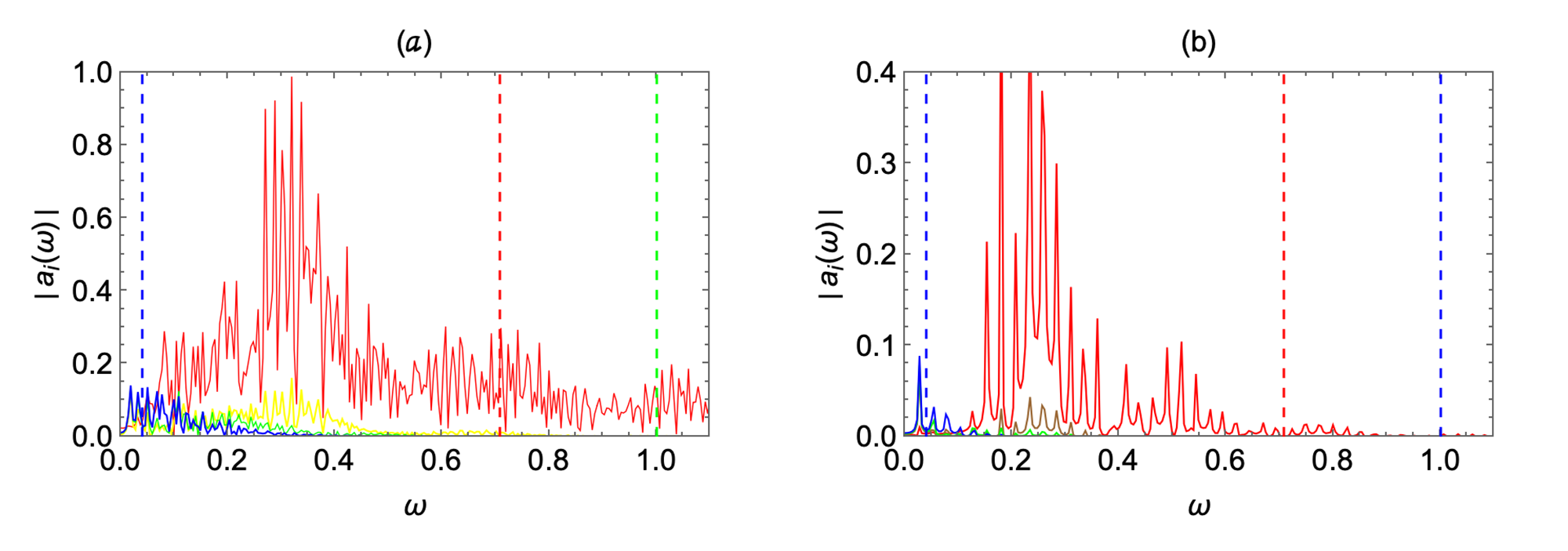}
\caption{ Acceleration spectra for the sub-kinks:  ${\bar k}_{-2}$   (red),   ${\bar k}_{-1}$  (yellow),   ${\bar k}_{1}$ (green),  and  ${\bar k}_{2}$  (blue), in a $K_4-{\bar K}_4$ collision, and its corresponding $m_i$ radiation thresholds  (vertical lines). $(a)$  Evaluated at $t= 2000$, just after the first rebound.  $(b)$ At   $t= 10000$, when the system has reached
a quasi-stable stage. The parameters are selected as: $n=4$, $\alpha=0.05$,  and $v_{in}=-0.1$.
  }
\label{Figallacel}
\end{center}
\end{figure}

\begin{figure}
 \begin{center}
\includegraphics[width=16.0cm,height=5.5cm]{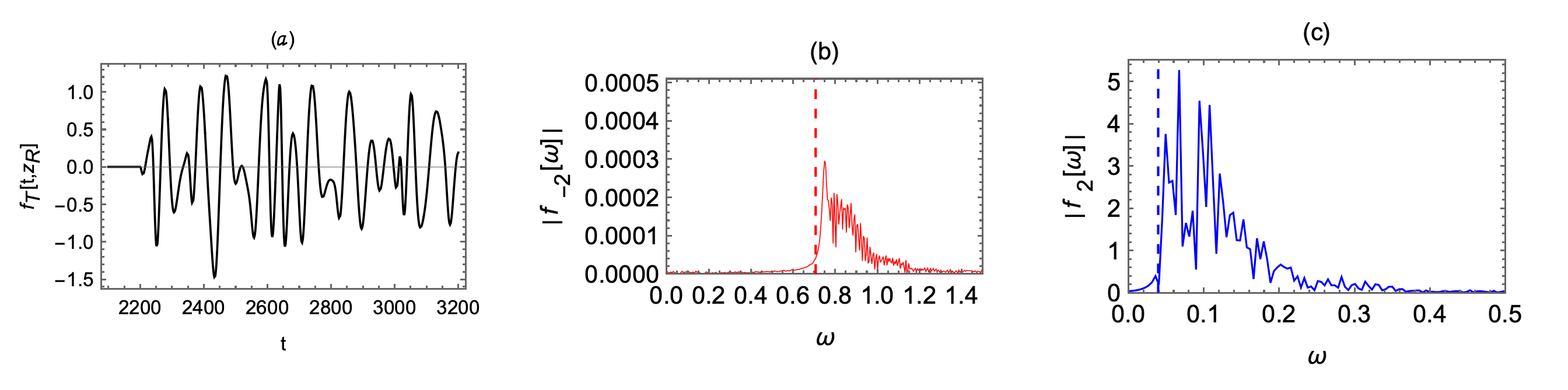}
\caption{ $(a)$ Total field radiation $f_T(t,z_R)$ obtained from the sum of the sub-kink contributions in  eq.~(\ref{eqrad3}).
 Field  $f_i(\omega,z_R)$ spectra for the:  $(b)$ $k_{-2}$ (red),   and $(c)$ $k_{2}$ (blue) sub-kinks,
  evaluated  after the first rebound,  $t= 2000$,  
and its corresponding $m_i$ radiation thresholds  (vertical lines). 
 The parameters are selected as $n=4$, $\alpha=0.05$,  $v_{in}=-0.1$, and  $ z_r=400$.
  }
\label{Figpf34}
\end{center}
\end{figure}

Figure~\ref{Figallacel} $(a)$ shows the acceleration spectra  $a_i(\omega)$, evaluated shortly after the first
rebound for each of the four sub-kinks, together with their respective radiation thresholds $m_i$.  As anticipated, the spectra contain a large set of resonant frequencies, yet only those associated with the external sub-kink lie above the radiation threshold. In contrast, figure~\ref{Figallacel} ($b$) shows that once the system settles into its quasi-stable stage, the acceleration spectrum is drastically reduced, exhibiting only a few dominant frequency peaks. Figure~\ref{Figpf34} displays the total radiation field $f_T(t,z_R) $ in the radiation region $z_R > x_i$, whereas the $f_i(\omega) $  Fourier spectra corroborates that  only the external sub-kink $k_2$ contributes significantly to radiation energy losses. 
 
For a higher incident speed, $v_{\text{in}} = -0.25$, the phase-space trajectories $(x_i,\dot{x}_i)$ shown in figure~\ref{FigPhaseS} ($b$) correspond to an inelastic collision in which the multi-kink rebounds with a final velocity $v_{\text{fin}} \simeq 0.16$.  In this case we have 
 $E_{CCM} (90000) -E_{CCM} (0) =
-0.39$ and $E_{rad} (0,90000) =  -0.43$.

\section{Multi-kink collisions}\label{MKAMKC}
 We study the collisions between two multi-kinks by numerically solving eq.~(\ref{eqphi}) with the initial ansatz
\ \begin{equation} 
\label{PhiKK*}
      \Phi_{K_4 \, { K^\prime_n}} (z)\, = \,    \Phi_{K_n}   \left( z+ L_n - v_{in} \, t  \right)  +   \Phi_{K_n}^{\prime}   \left( z- L_n + v_{in} \, t  \right) -n \pi   \, ,   
 \end{equation} 
which represents a head-on collision between two $n$ multi-kinks. The simulations employ a temporal step $\tau \approx 0.12$ and a spatial grid with spacing $h \approx 0.41$, comprising $12{,}000$ points over $z \in [-2500,2500]$. Absorbing boundaries are placed sufficiently far from the interaction region to prevent reflections. Initial multi-kink profiles are computed using a fourth-order Runge-Kutta scheme. We consider  initial separation  between the two mult-kinks  of the  order $L_n = 6 \, \times \sigma_{K_{n,\alpha}} $, where $\sigma_{K_{n,\alpha}} $ 
 is the multi-kink size given in eq.~(\ref{sigmaK2}), this choice guarantees negligible initial overlap between the colliding configurations. Throughout this section, the notation $K\{v_i,, v_i + 2n\pi\}$ is used to denote the vacuum sector in which a given multi-kink is defined.

 \begin{figure}
 \begin{center}
\includegraphics[width=10.1cm]{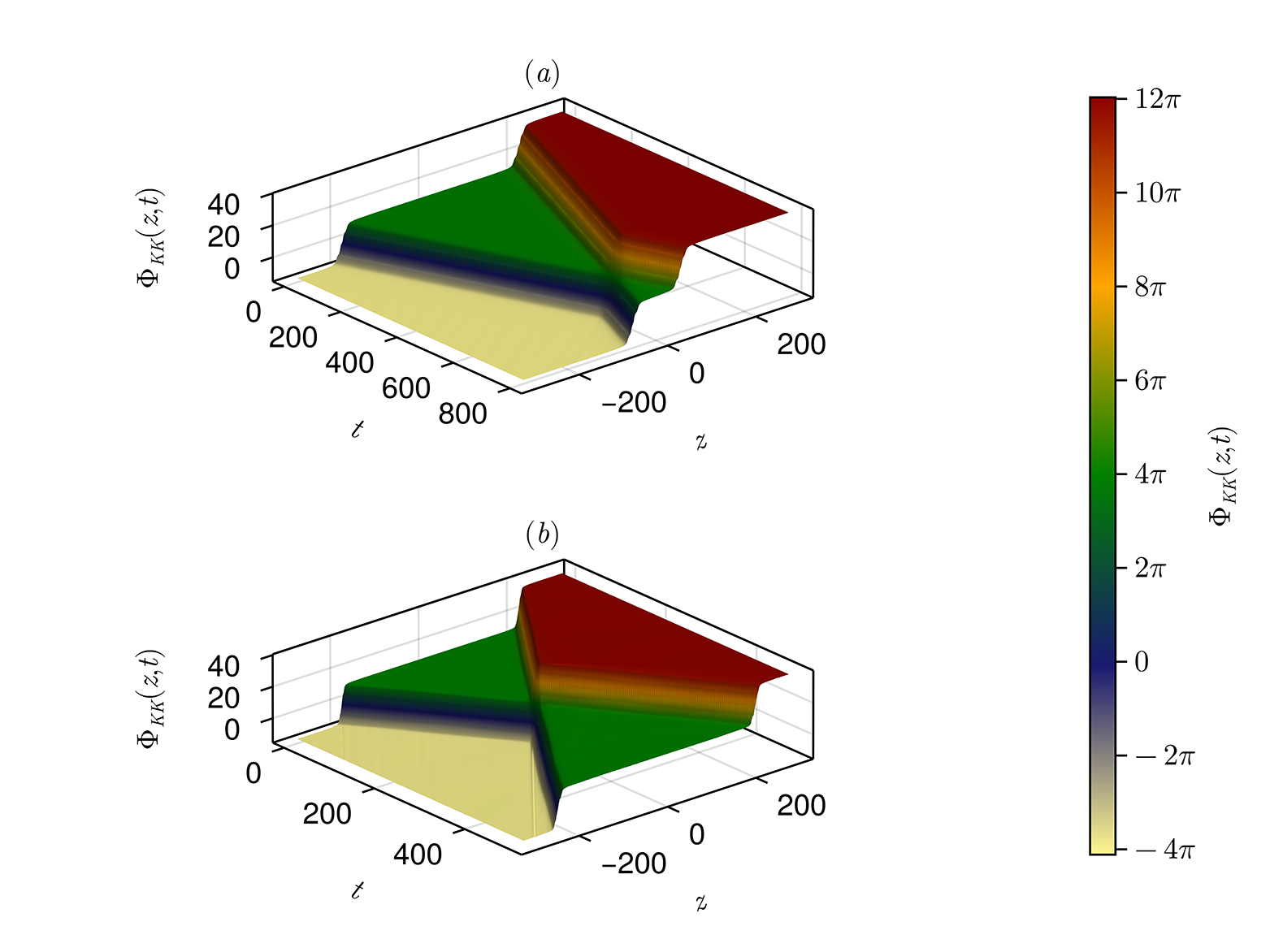}
\caption{ Space-time picture for the collision of two $n=4$ contiguous multi-kinks $K \{-4 \pi ,4 \pi  \} _{v_{in}} + K \{4 \pi ,12 \pi  \} _{-v_{in}}   \to K \{-4 \pi ,4 \pi  \} _{-v_{fi}}  + K \{4 \pi ,12 \pi  \} _{v_{fi}}  $.  The initial velocities are:  (a) 
$v_{in}=-0.2$ and  (b) $v_{in}=-0.9$, the other parameters are selected as  $L=200$ and $\alpha =0.05 $.
  }
\label{FigK4K4}
\end{center}
\end{figure}
 
 \subsection{$K_n- K_n$ collisions   }\label{KnKnsec}

We first examine the collision between two contiguous $n=4$ multi-kinks: $K\{-4\pi,4\pi \}_{v_{in}}   + K\{ 4\pi,12\pi \}_{-v_{in}} \Longrightarrow K\{-4\pi,4\pi \}_{-v_{fi}} + K\{ 4\pi,12\pi \}_{v_{fi}}$. The total topological charge $Q_{2\times4}=8$ is conserved, ensuring that the field always interpolates between $\phi \to -4\pi$ and $\phi \to 12\pi$ at spatial infinities.

As shown in figure~\ref{FigK4K4}, the internal structure of each multi-kink is largely preserved for a broad range of velocities, including $v_{\text{in}}=-0.2$ and $v_{\text{in}}=-0.9$. After the collision, both multi-kinks rebound and recover their initial profiles, moving apart with final velocities $v_{\text{fi}}$. The process resembles the elastic two-soliton collisions of the sine-Gordon model, though here it is  inelastic: small internal oscillations and weak radiation losses appear, yet the integrity of the multi-kink structure is preserved.

\begin{figure}
 \begin{center}
\includegraphics[width=16.0cm]{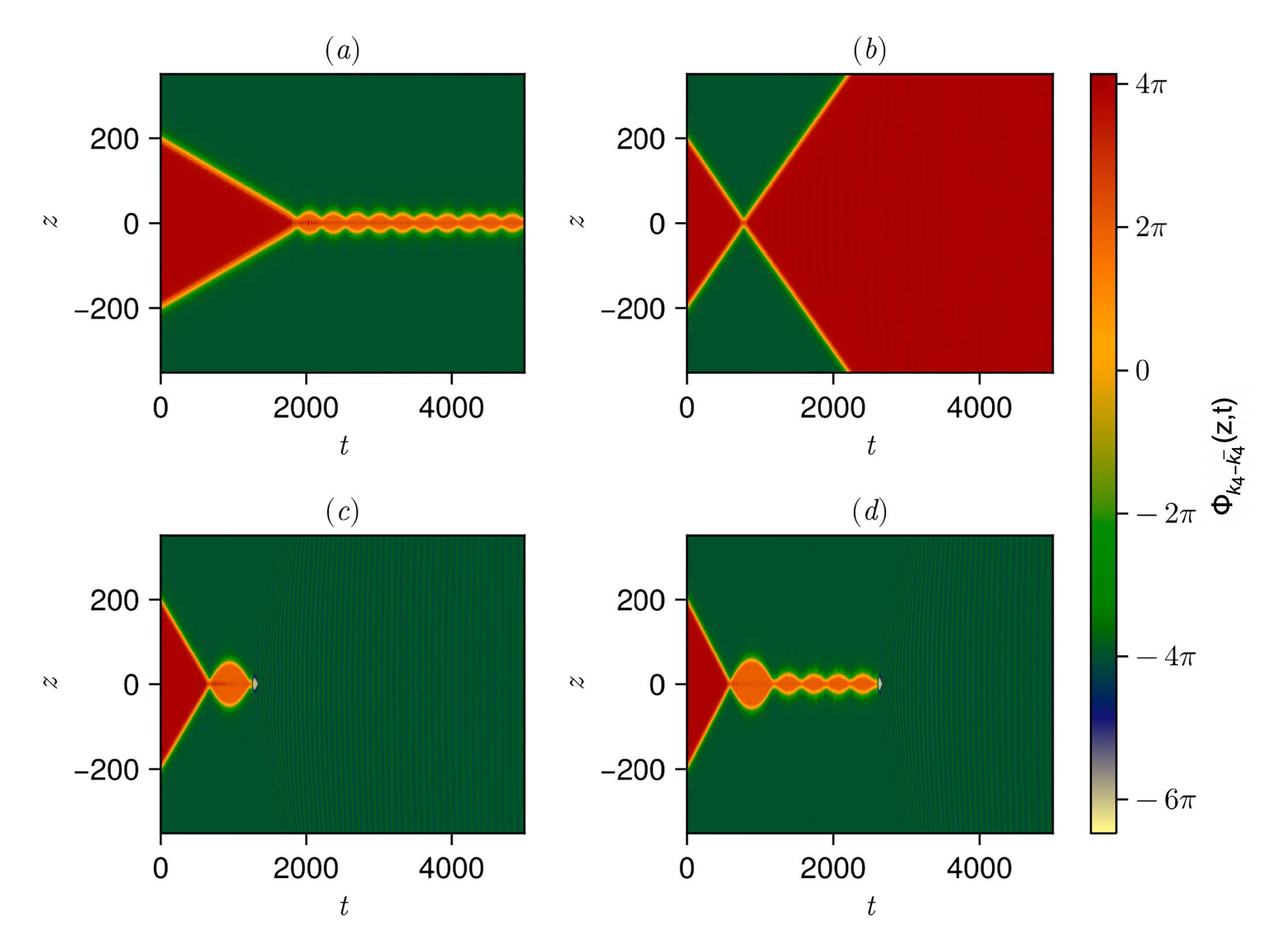}
\caption{ Space-time plots  of the    $K_4 - {\bar K}_4$ collisions 
 with various initial velocities: (a) $v_{in}=-0.1$,  (b)  $v_{in}=-0.25$, (c)  $v_{in}=-0.29$, and (d)  $v_{in}=-0.332$. In all cases $n=4$ and $\alpha=0.05$ and $L_4=200$.
  }
\label{FigK4AK4}
\end{center}
\end{figure}

 \subsection{ $K_n- {\bar K}_n$   collisions }\label{KnAKnsec}

 \subsubsection{  General results for the $n= 4,8,16$   cases }\label{KnAKnsec1}
    
We study the collision of a multi-kink  $K{\{-n\pi, n\pi\}}$ with its corresponding anti-multi-kink  $  {\bar K}{\{n\pi, -n\pi\}}$, considering the cases $n=4$, $8$, and $16$, which correspond to configurations containing different numbers of internal sub-kinks. For all values of $n$, the total topological charge of the combined $K_n-\bar K_n$ system is zero. As a consequence, the collision dynamics can lead to three qualitatively distinct outcomes, shown in figures~\ref{FigK4AK4}, \ref{FigK8AK8}, and \ref{FigK16AK16}:

\begin{figure}
 \begin{center}
\includegraphics[width=16.0cm]{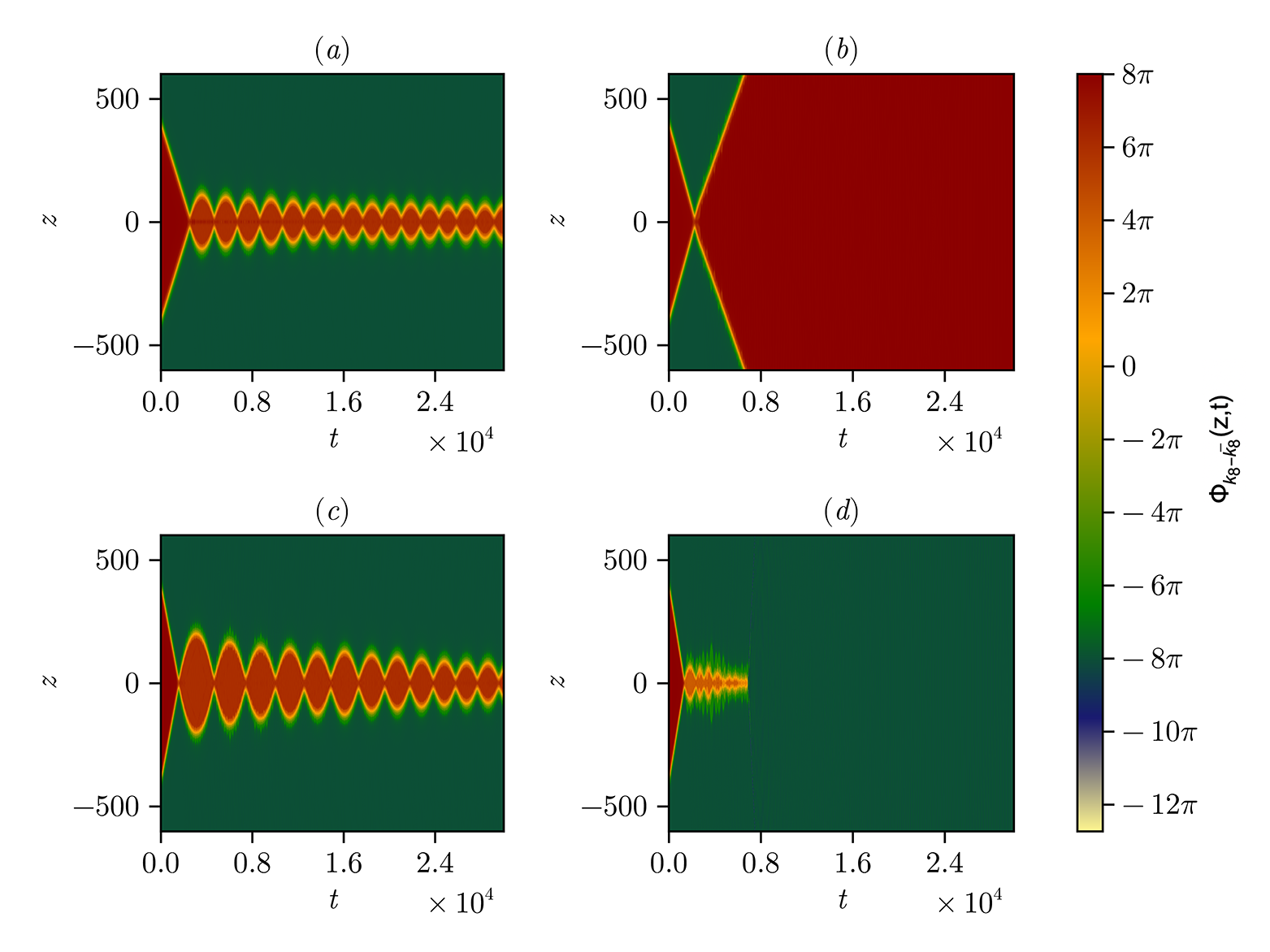}
\caption{ Space-time plots  of the    $K_8 - {\bar K}_8$ collisions 
 with various initial velocities: (a) $v_{in}=-0.15$,  (b)  $v_{in}=-0.171$, (c)  $v_{in}=-0.25$, and (d)  $v_{in}=-0.29$. In all cases  $n=8$, $\alpha=0.05$, and $L_8=600$,.
  }
\label{FigK8AK8}
\end{center}
\end{figure}

\begin{figure}
 \begin{center}
\includegraphics[width=16.0cm]{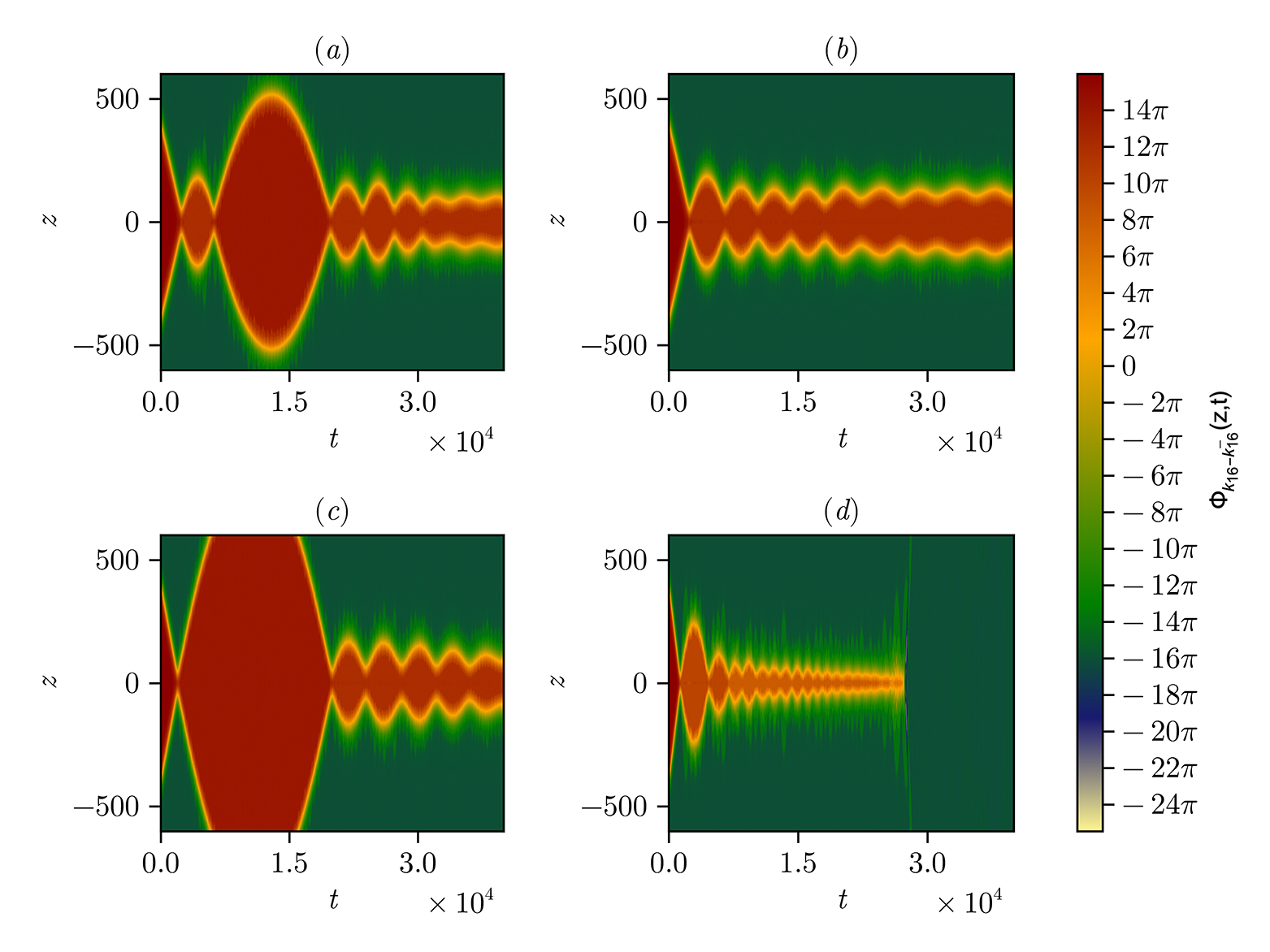}
\caption{ Space-time plots  of the    $K_{16} - {\bar K}_{16}$ collisions 
 with various initial velocities: (a) $v_{in}=-0.151$,  (b)  $v_{in}=-0.152$, (c)  $v_{in}=-0.19$, and (d)  $v_{in}=-0.29$. In all cases  $n=16$,  $\alpha=0.05$ and $L_{16}=1500$.
  }
\label{FigK16AK16}
\end{center}
\end{figure}

\emph{ (a) Oscillon formation}. In the low-velocity regime, $|v_{in}| < v_c \equiv v_c(n,\alpha)$, the collision leads to the formation of a long-lived bound state. This state can be identified as a composite oscillon that retains the internal multi-kink structure of the colliding solitons. Initially, the oscillon exhibits large-amplitude oscillations, which rapidly decrease before giving way to a stable regime characterized by an almost constant amplitude that persists for a very long time.

Once this regime is reached, we find that the oscillon size $R_{n,\alpha}$ is approximately given by $R_{n,\alpha} \approx  2 \, \sigma_{n,\alpha}$, where $\sigma_{n,\alpha}$ is the multi-kink size defined in eq.~(\ref{sigmaK2}). Consequently, the oscillon size follows the scaling behavior $R_{n,\alpha}   \sim   \frac{2n}{\pi} \, \ln(n)  \, \ln(\frac{1}{\alpha})$, which is corroborated by the numerical results shown in figures~\ref{FigK4AK4} $(a)$, \ref{FigK8AK8} $(a,c)$, and \ref{FigK16AK16} $(a,b,c)$.

 The oscillon configuration is characterized by a long-period envelope oscillation whose duration increases with $n$. This behavior is explained by the results presented in appendix~\ref{KnAKnsec}, where it is shown that the stable stage is mainly characterized by two oscillation frequencies: (i) a low frequency $\omega_L$, which lies below the scalar mass threshold, $\omega_L < m_{\frac{n}{2}}$; and (ii) a high frequency $\omega_H > m_{\frac{n}{2}}$, associated with linear radiation losses. According to eq.~\ref{wL}, the long-period oscillation is approximately estimated as $T_L = 2\pi/\omega_L \propto \sqrt{n/\alpha}$, which explains the observed behavior.

It is worth pointing out that, for $n=16$, the reduction of the oscillon amplitude toward the steady stage is not smooth. Instead, it exhibits the formation of intermittent large-amplitude bubbles, as shown in figure~\ref{FigK16AK16} $(a,c)$. This suggests that an ongoing bounce-dispersion process is interrupted by the strong $K_{16}-{\bar K}_{16}$ binding energy, which captures the two multi-solitons again.
  
 The picture of the oscillon structure described above is further supported by figure~\ref{Figbion}, which presents a sequence of snapshots illustrating the time evolution of the bound state emerging from the $K_4$-${\bar K}4$ collision. The resulting configuration can be interpreted as a composite large-amplitude oscillon: its outer regions retain the structure of the original multi-kinks, while the central region originates from the fusion of the innermost $k_2$-${\bar k}_{-2}$ sub-kinks, forming a localized bion that oscillates around the false vacuum at $v=(n-2)\pi$. This bion remains connected to the outer portions of the oscillon through the binding force given in eq.~(\ref{Force}).

\emph{ (b) Inelastic rebound.} When the impact velocity exceeds the critical threshold, $|v_{\rm in}|>v_c(n,\alpha)$, the collision generally leads to an inelastic dispersion inmediatly after the first bounce. Unlike the familiar behavior observed in the $\phi^4$, $\phi^6$, and related models \cite{Moshir1981,Campbell1983,Dorey2011,Dorey2021,Demirkaya2017,Moradi2017}, no multi-bounce escape windows have been found, in the extensive set of full-field simulations performed so far across the $(n,\alpha,v_{\rm in})$ parameter space.

\emph{ (c) Sudden  annihilations}.   In general, we find that annihilation eventually occurs whenever the parameter $\alpha$ exceeds a threshold value, $\alpha > \alpha_{trh}$. However, as shown in figures~\ref{FigK4AK4} $(c,d)$, abrupt annihilations can also take place at relatively short times for certain combinations of the parameters $\alpha$ and $v_{in}$. These observations suggest the existence of a non-uniform, and possibly chaotic, mechanism that triggers the sudden disappearance of the $K_n-{\bar K}_n$ configuration. Such behavior is not expected to arise from the linear radiation losses described in section~\ref{radsec}. Instead, as will be discussed in section~\ref{KnAKnsec3}, it results from a sequence of consecutive staccato bursts of radiation that rapidly drain energy from the system.

The emergence of the  three final states  previously described depend on the parameters $\alpha$ and $v_{\text{in}}$. This is illustrated  by the  $\alpha$-$v_{\text{in}}$   plot of $\Phi_{K_4 \bar{K}_4}(0,T)$. For $\alpha < \alpha_c \simeq 0.17$, low-velocity collisions tend to produce long-lived bound states, whereas higher incoming velocities lead to either dispersion or annihilation. In contrast, for $\alpha >\alpha_c$ the dynamics are dominated by annihilation events. The boundaries separating these regimes exhibit a highly intricate, fractal-like structure, characterized by transition layers and isolated islands.

\begin{figure}
 \begin{center}
\includegraphics[width=10.0cm]{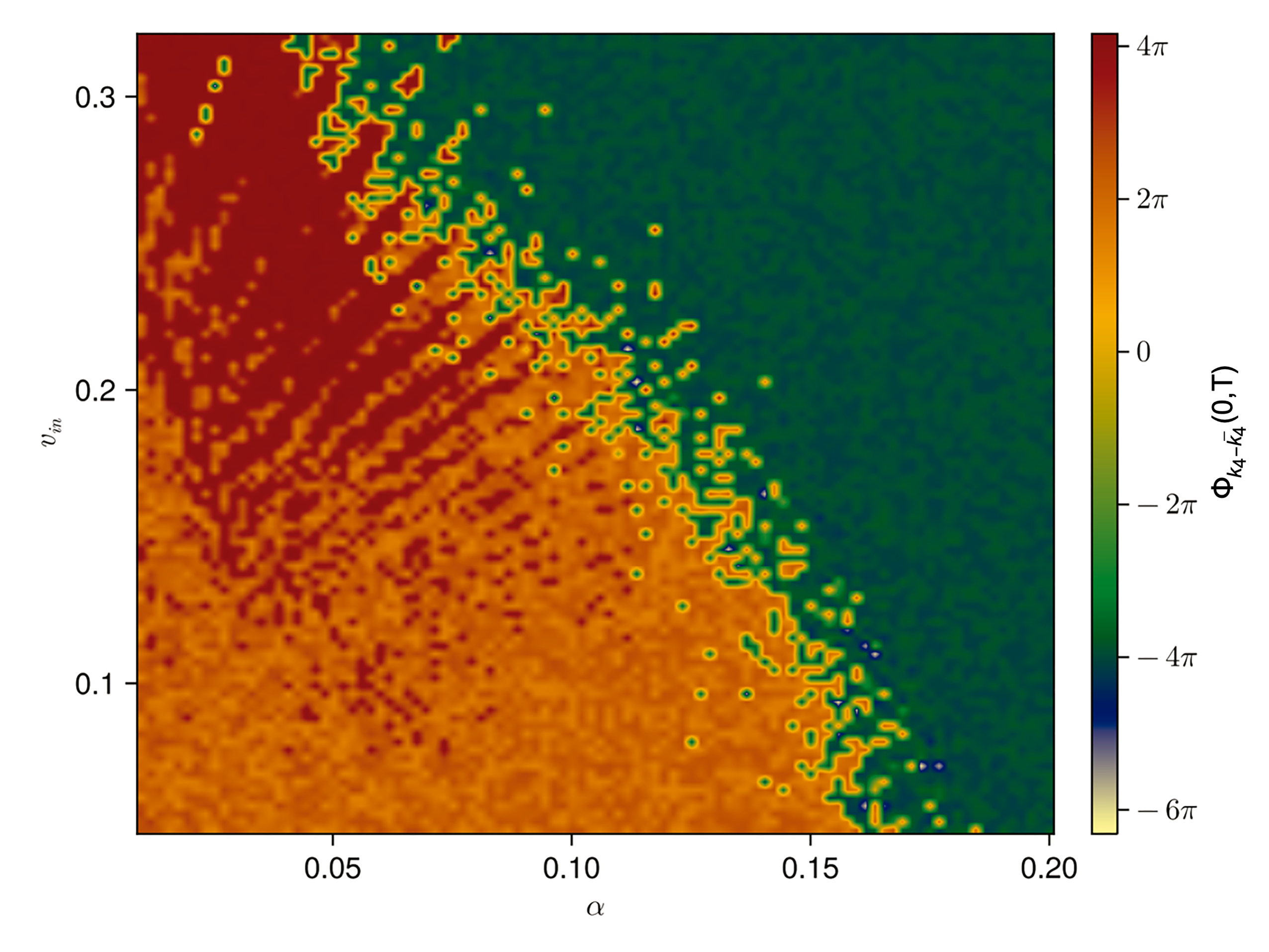}
\caption{ Field configuration $\Phi_{K \bar{K}}(0,T)$  for the  $K_4 -{ \bar K}_4$ collision, evaluated at the  centre of the   collision ($z=0$)   at a large  time  $T \approx 3000$, as a function of the  initial  speed $\vert v_{in} \vert $ and  the $\alpha$ breaking degeneracy parameter.  $T$ is measured once  the initial trajectories intersect.
  }
\label{FigalphviCC}
\end{center}
\end{figure}  

\begin{figure}
 \begin{center}
\includegraphics[width=12.0cm]{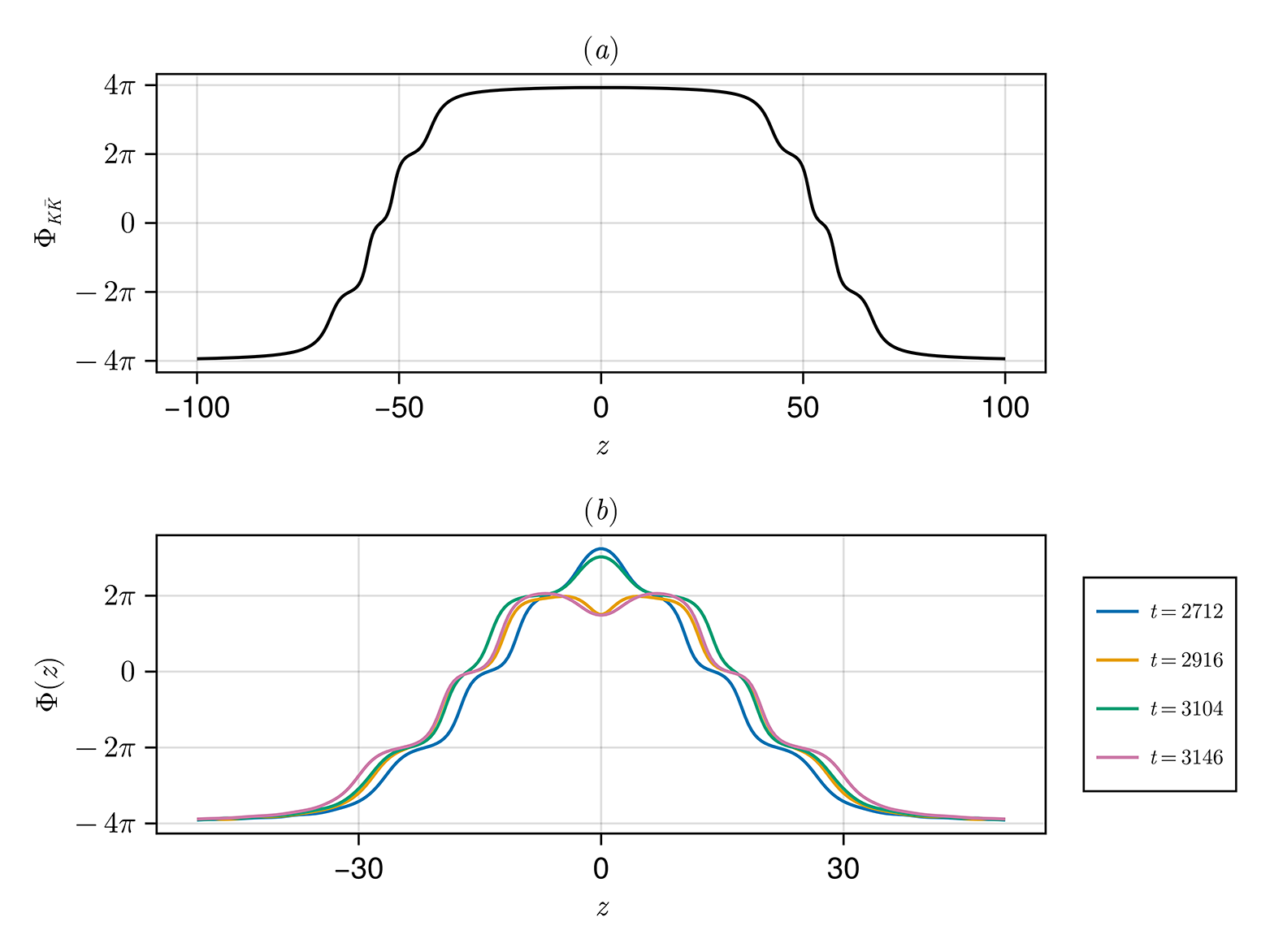}
\caption{ (a)    Field configuration  before  the    $K_4 -{ \bar K}_4$ collision. (b)  Post-collision field snapshots  at various times. The configuration can be identified with a  composite oscillon  that maintains part of  the colliding  multi-kinks structure, but its  central region corresponds to a  bion  that results from the merging of  $k_2$ and  ${\bar k}_{-2}$ sub-kinks.  The bion oscillates around the $\phi= 2\pi$ false vacuum.
The parameters  values are $\alpha =0.05 $, $n=4$, $L=200$ and  $v_{in}=-0.1$. 
  }
\label{Figbion}
\end{center}
\end{figure}

 \subsubsection{ Full field theory versus collective coordinate simulations }\label{KnAKnsec2}

We now compare the solutions of the full field equation~(\ref{eqphi}) with those obtained from the collective coordinate model, eqs.~(\ref{ccm3}), using equivalent initial conditions. To do so, we examine the time evolution of the coordinates $x_i(t)$ and contrast them with the positions of the energy density peaks, $X_{{\cal E}_i}(t)$, associated with each sub-kink. 

To determine  \(X_{\mathcal{E}_i} (t)\), we employ a time--slicing procedure. At each simulation time step \(t_s\), the energy density \(\mathcal{E}(t_s,z) \) is extracted from the numerical \(K_4-\bar{K}_4\) collision data. We then identify  the  \(\chi_i\)  localizations of   the local maxima of \(\mathcal{E}(t_s,\chi_i ) \) at fixed  \(t_s\), which sets  \(X_{\mathcal{E}_i} (t_s) \equiv \chi_i\).

 \begin{figure}
 \begin{center}
\includegraphics[width=15.0cm,height=10.0cm]{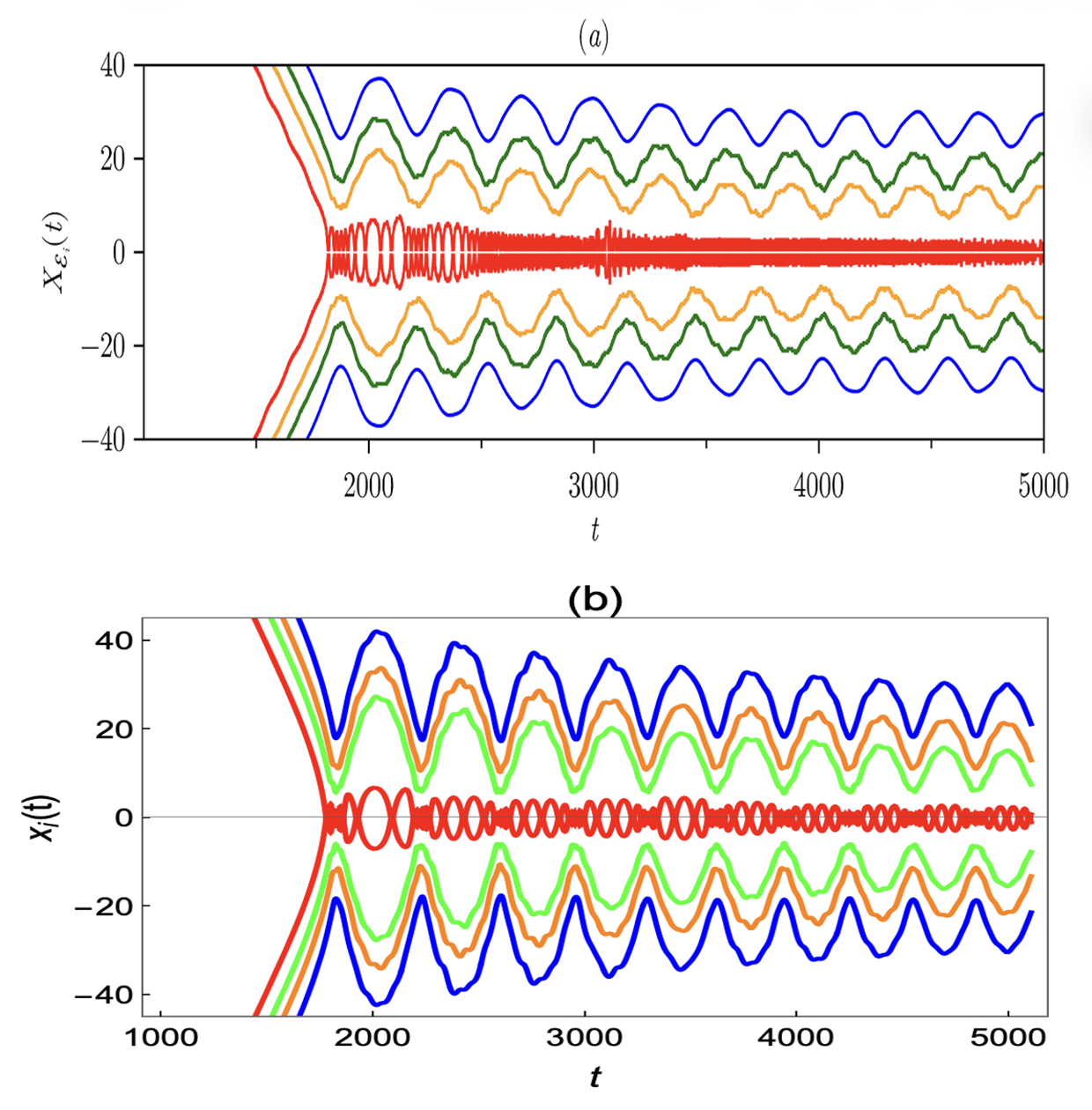}
\caption{ Oscillon evolution shortly after the   $K_4 -{ \bar K}_4$ collision, according the: $(a)$ Full field  or  $(b) $   collective coordinates descriptions.    The parameters  values are $\alpha =0.05 $, $n=4$, $L=200$ and  $v_{in}=-0.1$. 
  }
\label{FigComp-Osc-5}
\end{center}
\end{figure}


Figure~\ref{FigComp-Osc-5} shows the coordinates  evolution of an oscillon emerging from  a $K_4$-$\bar{K}_4$ collision. The collective coordinate approximation reproduces remarkably well the overall dynamical pattern observed in the full field results. In particular, the amplitud and long oscillation period of the  sub-kinks closely match those obtained from the field equations of motion. The most noticeable discrepancy arises in the  oscillation pattern of the central bion: while the  field results exhibits a sharp frequency increase after the third bounce, this effect appears less pronounced in the collective coordinate model.  

\begin{figure}
 \begin{center}
\includegraphics[width=14.0cm,height=10.0cm]{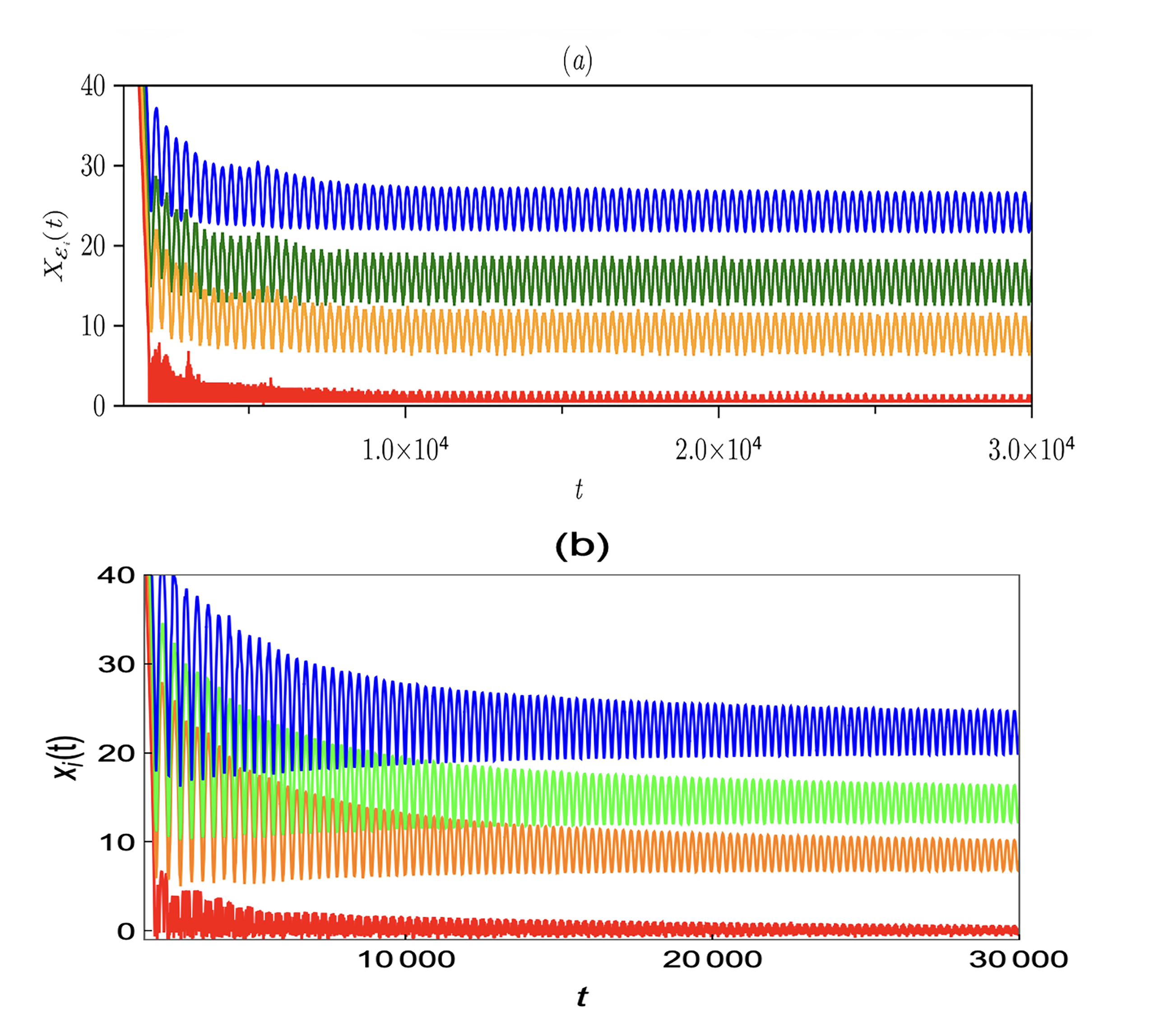}
\caption{ Oscillon evolution over a extended time period that includes the stable stage, according the: $(a)$ Full field  or  $(b) $   collective coordinates descriptions. Only positive values of the coordinates are displayed.  The parameters  values are $\alpha =0.05 $, $n=4$, $L=200$ and  $v_{in}=-0.1$.
  }
\label{Fig2-Osci-30}
\end{center}
\end{figure}

\begin{figure}
 \begin{center}
\includegraphics[width=12.0cm,height=10.0cm]{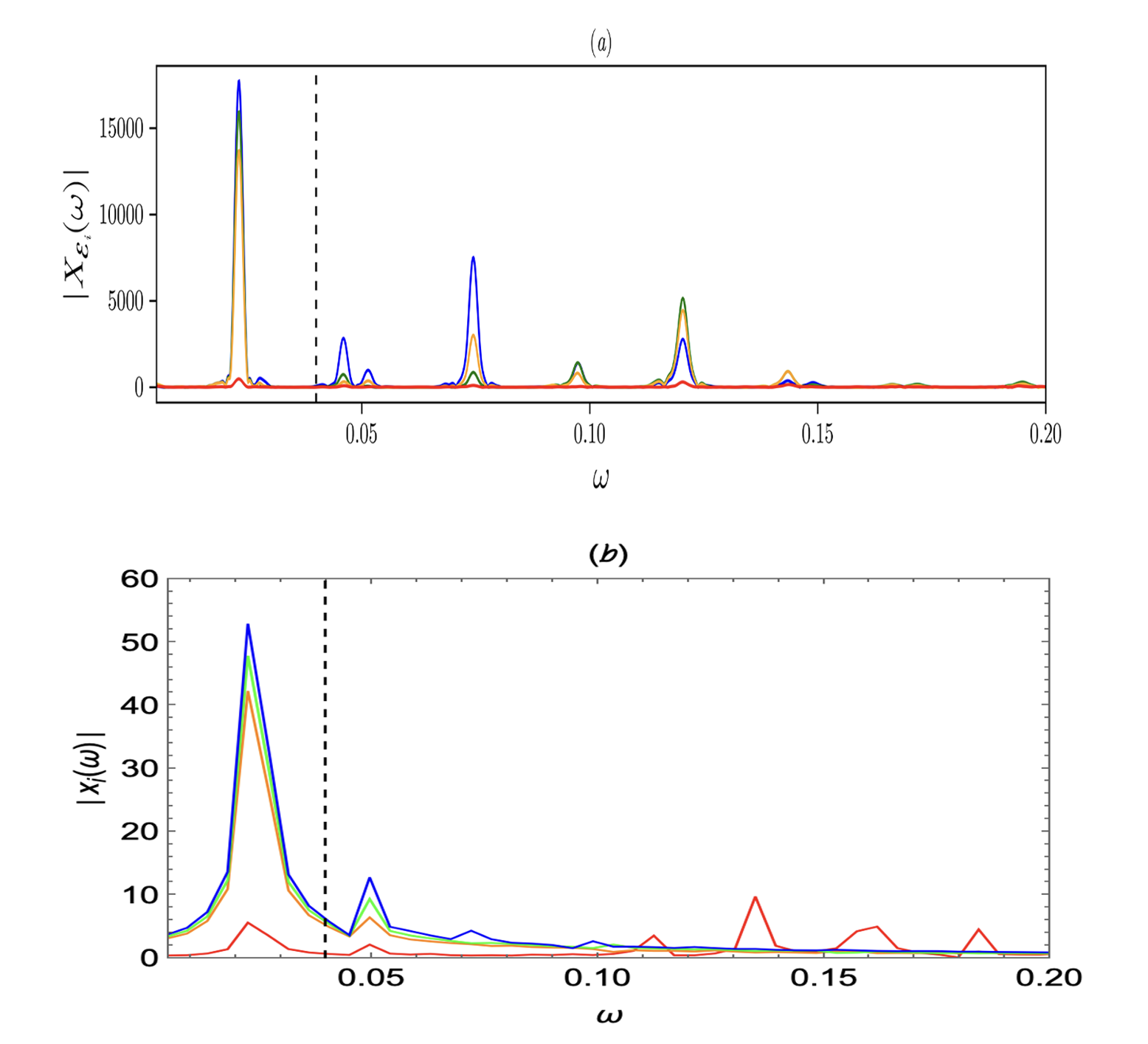}
\caption{ Fourier  frequency spectra for the $X_{{\cal E}_i} (\omega)$ and collective   $x_i (\omega)$ coordinates, evaluated at later time at which the stable stage has been achieved  $t= 100000$. The parameters selection is $\alpha =0.05 $, $n=4$, $L=200$ and $\vert v_{in} \vert=0.1$. 
  }
\label{Figxiwt}
\end{center}
\end{figure}

To investigate the long-time behavior of the oscillon, figure~\ref{Fig2-Osci-30} shows the evolution of $x_i(t)$ and $X_{{\cal E}i}(t)$ over an extended time interval. Once again, excellent agreement between the two approaches is observed. A characteristic feature of the dynamics is the abrupt reduction of the oscillon amplitude after the first few bounces, from which the system settles into a long-lived, nearly stable configuration that radiates only a small amount of energy. During this stage, the oscillon preserves both its shape and internal structure while propagating, exhibiting a remarkable degree of quasi-periodicity. This behavior is evident in the Fourier spectra of the coordinates, shown in figure~\ref{Figxiwt}, which display a set of narrow spectral peaks. The dominant peak corresponds to a low-frequency mode, $\omega_L$, located well below all relevant mass thresholds, see also the appendix~\ref{KnAKnsec}. Only a few small peaks associated with the outer sub-kinks extend above the $m_{\frac{n}{2}}$ mass threshold, indicating that radiative losses are weak and are emitted primarily by the external sub-kinks. The higher-frequency components correspond to overtones of the dominant mode, satisfying $\omega_j \approx j,\omega_L$, with $j=1,2,\ldots,5$. The agreement between the $X_{{\cal E}_i}(\omega)$ and $x_i(\omega)$ spectra is clearly visible for the low-frequency modes, although it becomes less pronounced for the higher harmonics.

We highlight the emergence of a synchronization phenomenon, shown in  figure~\ref{Figxiwt}, in which the sub-kinks adjust their internal oscillation modes and exhibit a coherent collective dynamics. This synchronization causes them to oscillate at shared frequencies, as evidenced by several of the resonant peaks in the spectra.

The results collected in Table~\ref{table1} reveal the impact of varying $\alpha$ on the oscillon properties. In particular, the large oscillation period $T_L$ decreases as $\alpha$ increases, with reasonably good agreement between the collective-coordinate predictions and the full field-theory results. Fitting the field-theory data yields the empirical relation $ T_L \approx  \, 80 \, \alpha^{-0.4}.
$ This scaling can be compared with the prediction obtained from the small-oscillation approximation, eq.~(\ref{wL})
$ T_L \approx  \, 53 \, \alpha^{-0.5}. $ Although the latter does not provide a quantitatively accurate estimate of the period, it correctly reproduces the overall trend and predicts a similar dependence on $\alpha$.

\begin{figure}
 \begin{center}
\includegraphics[width=12.0cm,height=8.0cm]{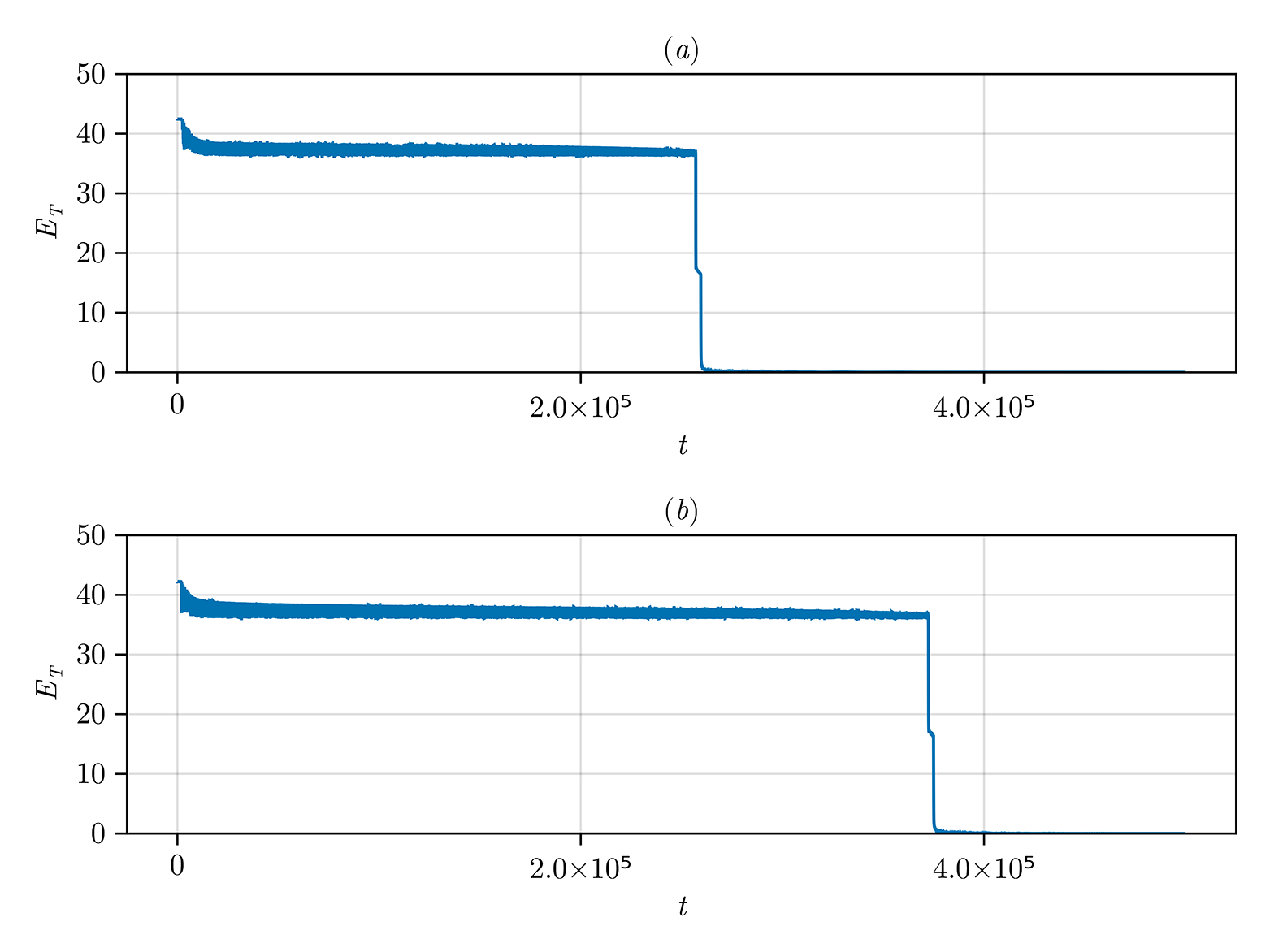}
\caption{ Total energy oscillon decay $E_T(t)$ for $(a)$ $\alpha =0.05 $ and  $(b)$ $\alpha =0.04 $. The other parameters selection is  $n=4$, $L=200$ and  $\vert v_{in} \vert=0.1$. 
  }
\label{FigLong-Live-O}
\end{center}
\end{figure}

\subsubsection{ Oscillon's decay and  $K_n-{\bar K}_n$ annihilation   }\label{KnAKnsec3}

To assess the oscillon's longevity, we present in figure~\ref{FigLong-Live-O} the evolution of the oscillon energy as a function of time. The results show that the oscillon maintains a remarkably large and nearly constant amplitude over an extended period before collapsing abruptly. Taking into account the results summarized in table~\ref{table1}, we find that the oscillon lifetime can reach values as large as
$\tau_{osc} \sim 5 \times 10^7 \times m_0^{-1} $, 
and increases as $\alpha$ decreases, following an empirical scaling relation
$\tau_{osc} \sim 4.8 \times \alpha^{-3.5} \times m_0^{-1} $, that is obtained fitting the field-theory data on  table~\ref{table1}.
These findings suggest that, in principle, large size oscillons of arbitrarily  long lifetimes may be obtained for sufficiently small $\alpha$.

We note that this sudden oscillon collapse cannot be explained by the linear radiation mechanism considered previously. That mechanism leads instead to a smooth, gradual energy loss, characterized by an approximately exponential decay associated with the oscillations of the external sub-kinks around the true vacuum. In contrast, the abrupt disappearance of the oscillon suggests the presence of an additional instability mechanism that is not captured by the linear radiation analysis.

\begin{figure}
 \begin{center}
\includegraphics[width=12.0cm,height=8.0cm]{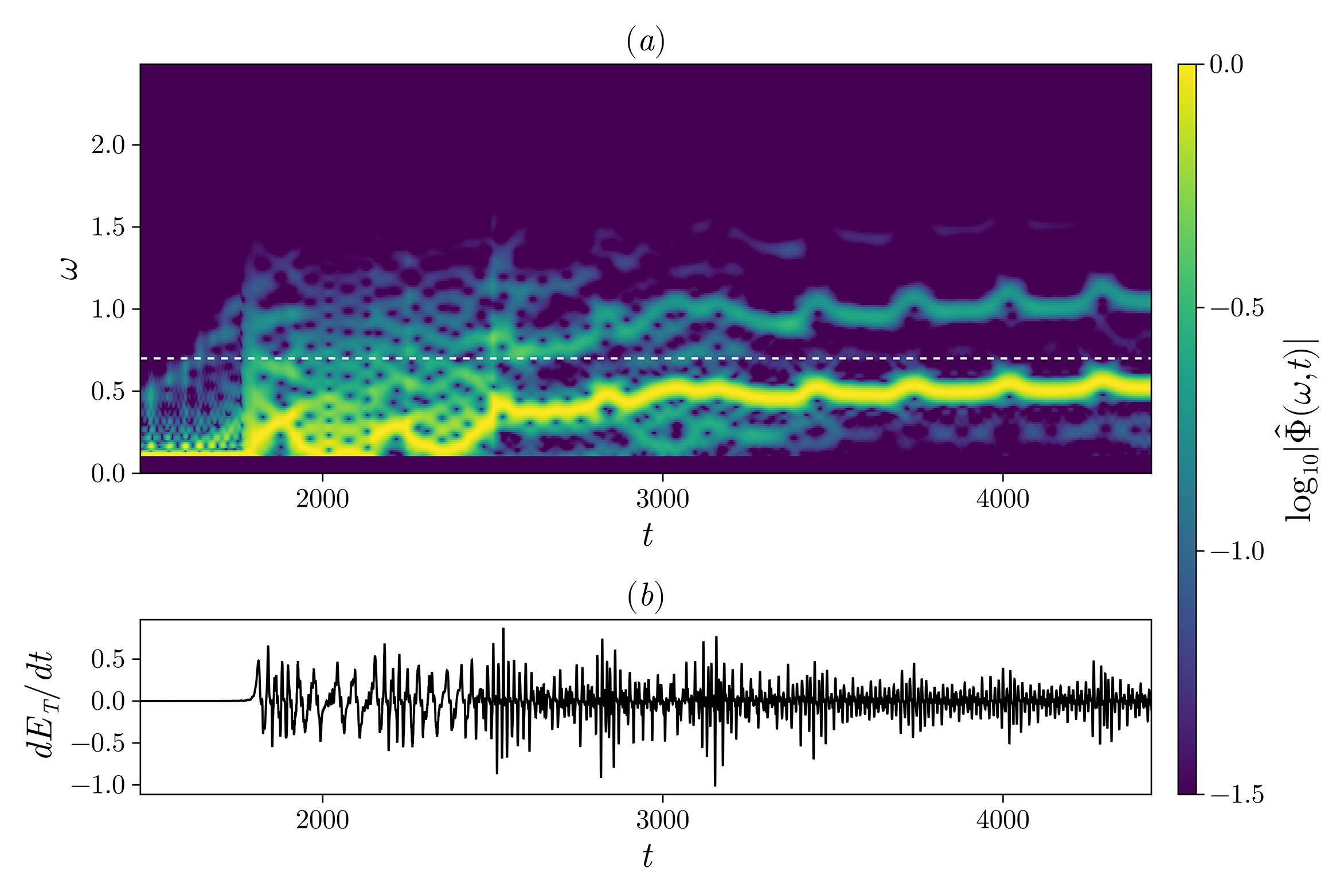}
\caption{ 
$K_4-{\bar K}_4$ collision leading to an oscillon formation,  for  $n=4$, $\alpha=0.05$ and $ \vert v_{in} \vert =0.1$.  The mass radiation threshold $ m_1 =0.71$ is represented by the white-dashed line. (a) Spectrogram of the 
$\Phi_{K_4-{\bar K}4} \ (0,t)$ field at the centre $z=0$. (b) Differential radiation energy losses $dE_T/dt$ as a function of time. }
\label{KAK-Bion-spect}
\end{center}
\end{figure}

\begin{figure}
 \begin{center}
\includegraphics[width=12.0cm,height=8.0cm]{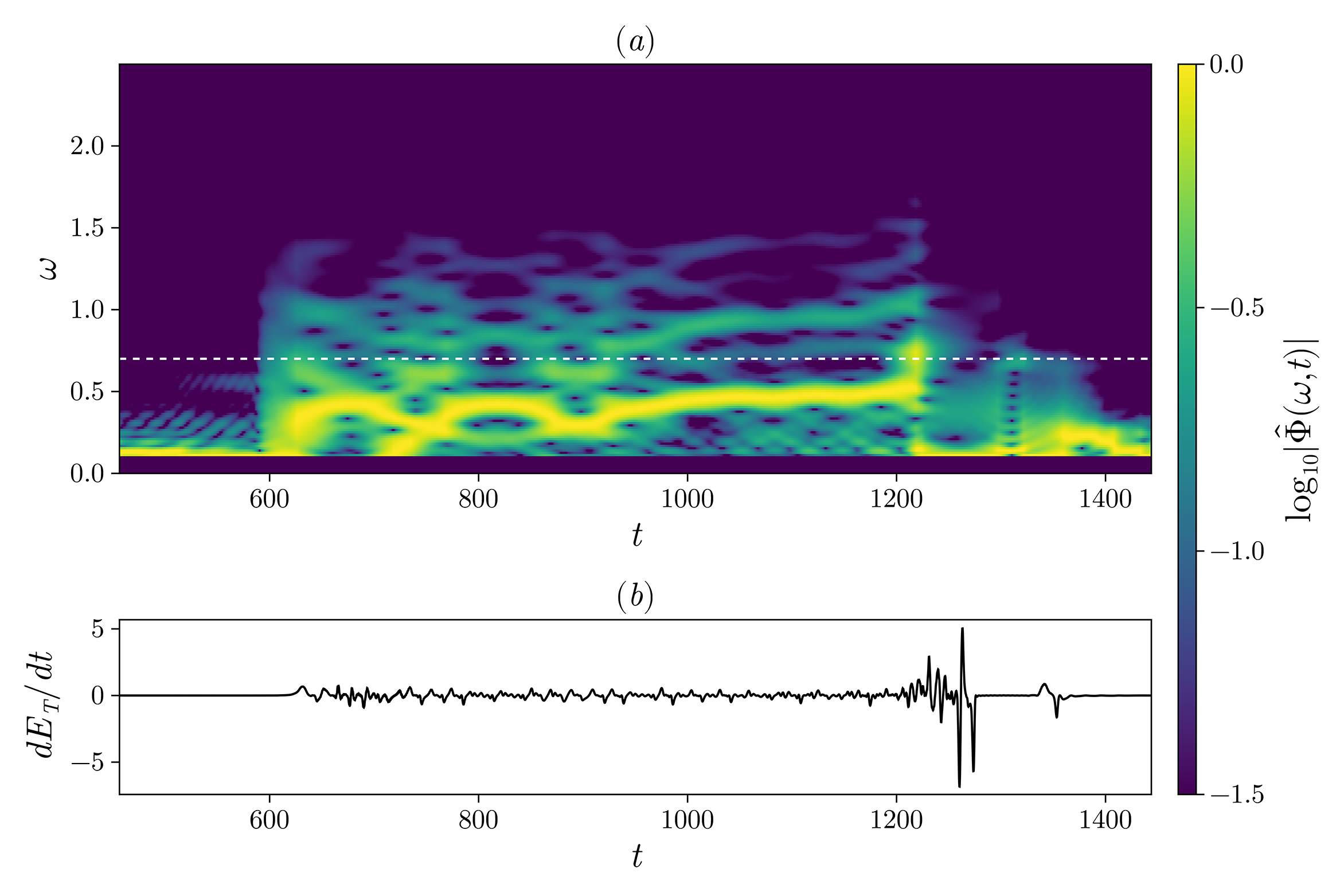}
\caption{ 
$K_4-{\bar K}_4$ collision leading to a total annihilation after the first bounce,  for  $n=4$, $\alpha=0.05$ and $ \vert v_{in} \vert =0.29$.  The mass radiation threshold $ m_1 =0.71$ is represented by the white-dashed line. (a) Spectrogram of the 
$\Phi_{K_4-{\bar K}4} \ (0,t)$ field at the centre $z=0$. (b) Differential radiation energy losses $dE_T/dt$ as a function of time. }
\label{KAK-Anh-2B-spect}
\end{center}
\end{figure}

\begin{figure}
 \begin{center}
\includegraphics[width=12.0cm,height=8.0cm]{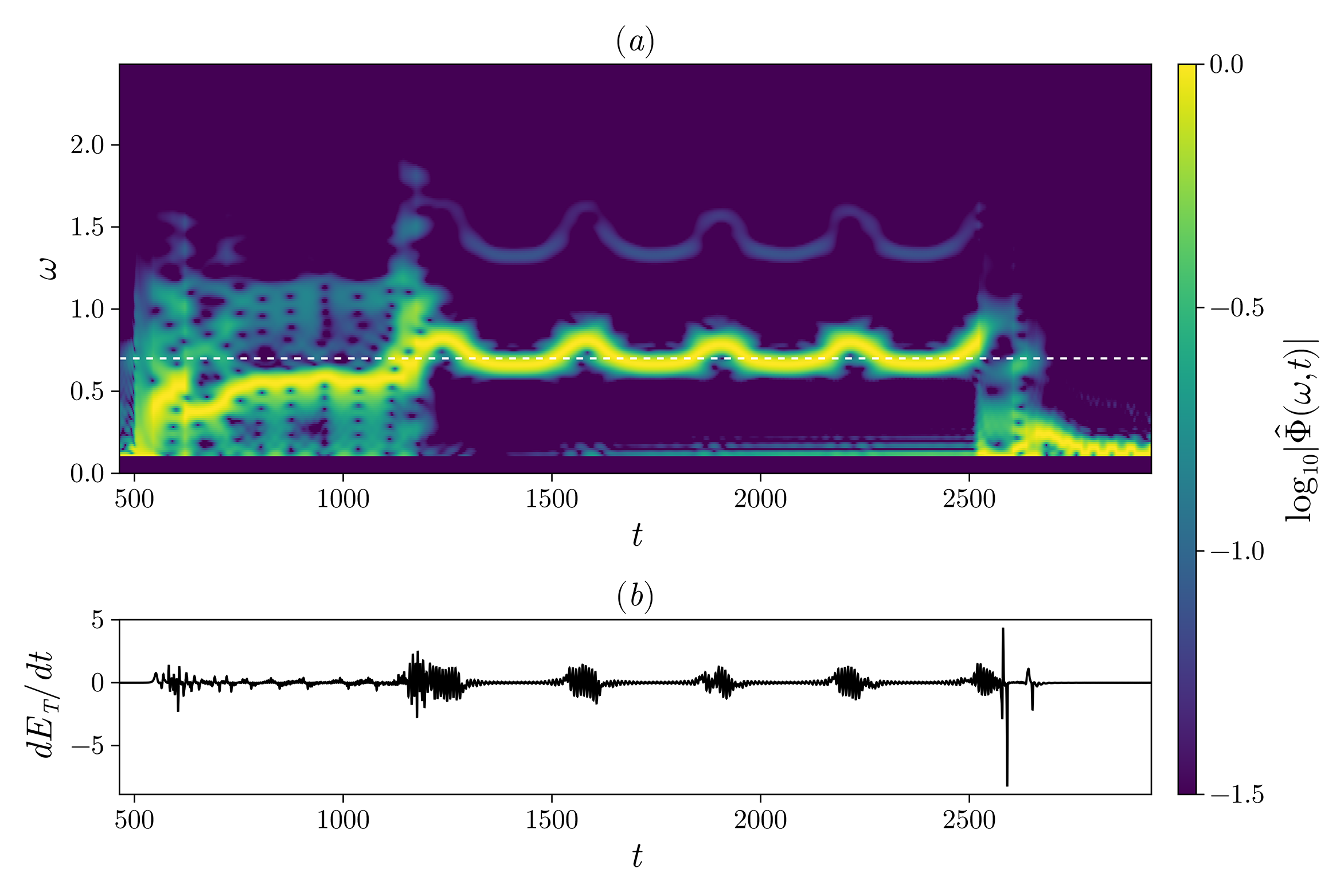}
\caption{ 
$K_4-{\bar K}_4$ collision leading to a total annihilation after  five bounces,  for  $n=4$, $\alpha=0.05$ and $ \vert v_{in} \vert =0.332$.  The mass radiation threshold $ m_1 =0.71$ is represented by the white-dashed line. (a) Spectrogram of the 
$\Phi_{K_4-{\bar K}4} \ (0,t)$ field at the centre $z=0$. (b) Differential radiation energy losses $dE_T/dt$ as a function of time. }
\label{KAK-Anh-5B-spect}
\end{center}
\end{figure}

It is therefore relevant to investigate the evolution of the bion that binds the two multi-kinks into a single oscillon. To this end, figures~\ref{KAK-Bion-spect}, \ref{KAK-Anh-2B-spect}, and \ref{KAK-Anh-5B-spect} display spectrograms of the central field signal, $\Phi_{K_4-{\bar K}_4}(z=0,t)$, for three representative $K_4-{\bar K}_4$ collisions shown in figure~\ref{FigK4AK4}.

In these spectrograms, the yellow regions correspond to the dominant frequency component of the signal. At early times, the dominant frequency $\omega_D$ is relatively small, with values of the order of $\omega_D \sim \omega_H \sim 0.08$--$0.22$ for $\alpha=0.05$, in agreement with eq.~(\ref{wfreq1}) and~(\ref{wH}). Notice that this fundamental low-frequency mode is accompanied by several weaker higher harmonics, visible as green bands in the spectrograms. The different collision outcomes can be summarized as follows:

(i) Oscillon formation when $ \vert v_{in} \vert =0.1$, figures~\ref{FigK4AK4} $(a)$ and \ref{KAK-Bion-spect}. It is worth noting that the rapid transition toward the stable stage described previously is accompanied by a slow, adiabatic increase of the fundamental frequency. During this process, some of the higher harmonics (green bands) cross the $m_1$ radiation threshold, leading to a small amount of energy being emitted. Eventually, however, the frequency growth ceases and the fundamental mode stabilizes just below the radiation threshold. The system then remains in this state for a very long time, effectively suppressing further radiative losses and thereby accounting for the oscillon's exceptionally long lifetime.

(ii) $K_4-{\bar K}_4$ annihilation after one bounce, figures~\ref{FigK4AK4} $(c)$ and \ref{KAK-Anh-2B-spect}. For a higher incoming velocity, $\vert v_i \vert = 0.229$, the dominant frequency increases until it crosses the radiation threshold. As shown in figure~\ref{KAK-Anh-2B-spect}, this crossing triggers a single, intense burst of radiation, releasing enough energy to destabilize the bion and cause its immediate annihilation.

(iii) $K_4-{\bar K}_4$ annihilation after five bounces, figures~\ref{FigK4AK4} $(d)$ and \ref{KAK-Anh-5B-spect}. Finally, for $\vert v_i \vert = 0.332$, a novel and remarkable phenomenology emerges. As shown in figure~\ref{KAK-Anh-5B-spect}, the dominant frequency initially crosses the radiation threshold, producing an intense burst of radiation. However, instead of continuing to higher frequencies, the mode subsequently reverses its evolution and recrosses the threshold multiple times. In total, five successive threshold crossings are observed, each accompanied by a new radiation burst. Remarkably, these bursts are not generated by different harmonics but rather by a single re-entrant dominant mode.

In the conventional picture, once a bound-state frequency crosses the mass threshold it transforms into a propagating mesonic excitation that escapes to infinity~\cite{Dorey2020}. In the present model, however, the bion appears to be partially confined by the outer sub-kink components, which may prevent the complete escape of the excitation and lead to the repeated threshold crossings observed in the spectrogram. 

In the previous figures, $dE/dt$ denotes the time derivative of the bion energy  $ E=\int_{-10}^{10}\mathcal{E}(z,t)dz$.  hence it measures the energy radiated directly from the central region. As shown in figure~\ref{KAK-Bion-spect}, this quantity remains approximately constant, on average, throughout the oscillon lifetime. Its magnitude, however, is negligible compared to the energy carried away by each of the staccato radiation bursts shown in figures~\ref{KAK-Anh-2B-spect} and \ref{KAK-Anh-5B-spect}. 
Therefore, we can  conclude that  these bursts, not only  constitute the dominant energy-loss mechanism responsible for the rapid annihilation  observed in some  $K_n-{\bar K}_n$ collisions, but also explain the final collapse of the oscillons.

Summarizing, the existence of long-lived oscillons requires the fulfillment of the following conditions:
\begin{equation} 
\label{conLLosc}
    \omega_L < m_{\frac{n}{2}} \,, \quad \quad m_{\frac{n}{2}} < \omega_H,   \quad  \quad \omega_H < m_{\frac{n}{2}-1} \, ,   
 \end{equation} 
where $\omega_L$ and $\omega_H$ can be estimated within the small-oscillation approximation using eqs.~(\ref{wfreq1}) and~(\ref{wH}). The first condition guarantees the stability of the long-period oscillations of the oscillon's external components. The second condition allows a moderate energy loss through radiation emitted by the external sub-kinks, thereby promoting the synchronization of the oscillon's vibrational components. The third condition suppresses the rapid decay of the central bion and prevents the collapse of the oscillon over long time scales, until the condition eventually ceases to be satisfied. The precise mechanism responsible for these features, and its connection with the eventual bion annihilation and oscillon collapse, remains an open question that warrants further investigation.

\begin{table}
\begin{center}
\resizebox{10cm}{!} {
 \begin{tabular}{ p{2.0cm}p{2.0cm}p{2.0cm}p{2.0cm}p{2.0cm}  }
 \hline
 \multicolumn{5}{c}{ \bf  \large Oscillon long-periods   and life-times   } \\ 
 \cline{1-5}  
 & FF  & CCM &  FF & CCM \\
\hline
 \hline
$\alpha$  &  $ T_L$   & $T_L$   &   $\tau $ &   $\tau $ \\
 \hline
$0.01$  & 862    &  665  & $4.6 \times 10^7$  &  $1.8 \times 10^5$    \\

 $0.02$ &   556  &  460 & $4.1 \times 10^6$  &   $9.7 \times 10^4$  \\
 
$0.03$ & 346 &   360 &    $9.9 \times 10^5$  & $8.2 \times 10^4$ \\

 $0.04$    & 302 &   322 &  $3.8 \times 10^5$   &  $5.4 \times 10^4$  \\
 
$0.05$ &   272  &  275 &     $2.8 \times 10^5$  & $4.5 \times 10^4$  \\ 

 $0.06$ & 258  & 244 &   $6.1 \times 10^4$   &   $3.9 \times 10^4$ \\
 
 $0.07$ &  243 &  232 &   $ 3.5 \times 10^4$  &   $3.5 \times 10^4$    \\
 
 $0.08$    &230 &   218 &  $ 2.6 \times 10^4$    &    $3.1 \times 10^4$    \\
 
$0.09$ &   225  &  202 & $2.0 \times 10^4$   & $2.8 \times 10^4$     \\

 $0.1$ & 204 &  196 & $2.7 \times 10^4$ &   $2.5 \times 10^4$  \\
 \hline 
\end{tabular}
}
\caption{ Oscillon large periods $T_L$ and life-times  $\tau $ as a function of $\alpha$, obtained from the full field (FF) or collective coordinate (CCM) schemes. }
\label{table1}
 \end{center}
\end{table}


\section{Conclusions}\label{conclu}

The aim of this work was to construct a commensurable and non-degenerate double sine-Gordon theory that exhibits a rich phenomenology, enabling the exploration of novel multi-soliton and oscillon properties. In the degenerate limit $\alpha = 0$, the model supports $n$ kink solutions interpolating between consecutive absolute minima. In contrast, moving into the $\alpha > 0$ regime induces a partial lifting of the vacuum degeneracy, providing a natural mechanism for the emergence of static multi-kinks $K_n$ composed of an arbitrary number $n$ of internal sub-kinks. Although exact analytical field configurations could not be obtained, the smooth-modulation approximation performs remarkably well and allows for a detailed and comprehensive analysis of the multi-kink sector.

The multi-kink $K_n$ is characterized by a topological charge $Q_n = n$, equal to the sum of the individual sub-kink charges, and by a mass $M_{K(n,\alpha)}$ eqs.~(\ref{MKSM1},\ref{MKSM2}), which results from adding the sub-kink masses eq.~(\ref{SumaMi}) together with a binding-energy contribution $\Delta_{K(n,\alpha)}$ eq.~(\ref{Vint}). In the large-$n$ limit, all these quantities scale linearly with $n$. The multi-kink structure, with its well-defined internal energy lumps, becomes particularly clear in the small-$\alpha$ regime, where we identified the following scaling relations:
(a) the equilibrium distance between adjacent sub-kinks obeys $  \xi_{i+1} - \xi_{i}    \approx   \frac{1}{m_{i}}   \log \left( \frac{const}{\alpha} \right)$  eqs.~(\ref{XeqG},\ref{XeqSM2});
 (b) the total size satisfies   $\sigma_{K_{(n,\alpha)}} \approx   \frac{2n}{\pi} \, \ln(n)  \, \ln(\frac{32}{\alpha}) $ eq.~(\ref{sigmaK2}); and (c) the binding energy scales as $\Delta V_{K_{(n,\alpha)}} \approx 
\frac{2n}{\pi} \, \alpha \,  \, \ln(\frac{12}{\alpha})$ eq.~(\ref{MKSM3}). Thereby, we can visualize the multi-kink as a large size,   weakly bound n-soliton.

For the $K_n- {\bar K}_n$ scattering, we identified three distinct dynamical regimes:
(a) Rapid annihilation, occurring after only a few bounces;
(b) Inelastic dispersion, in which the colliding configurations bounce back and separate to infinity. Unlike the familiar behavior of the $\phi^4$, $\phi^6$, and related models \cite{Moshir1981,Campbell1983,Dorey2011,Dorey2021,Demirkaya2017,Moradi2017}, no multi-bounce escape windows were observed. 
(c) The formation of long-lived compound oscillons, which constitutes the most novel outcome. These composite oscillons reflect the internal structure of the parent multi-kinks and therefore possess multiple characteristic scalar masses eq.~(\ref{msc}) and  resonant oscillation frequencies eqs.~(\ref{w1p1},\ref{wKAK}). These features distinguishes them from most oscillons reported in the literature.

We also developed a collective coordinate model that offers a good approximation to the full field dynamics during multi-kink collisions. Its construction involves several key elements: the use of dynamical Bogomol'nyi relations to determine the inter-sub-kink forces eq.~(\ref{Force}) and the $K_n - {\bar K}_n$ interaction force eq.~(\ref{FKaK2}); and the formulation of an approximate  linear radiation scheme that exploits the fact that the oscillation frequencies of the internal sub-kinks lie below the radiation threshold. As a result, radiation losses arise predominantly from the external sub-kinks, giving rise to an effective radiation-reaction force eq.~(\ref{eqrad6}). All these components are incorporated into the equations defining the collective coordinate model eq.~(\ref{ccm3}).

The field-theoretic solutions, together with those obtained from the collective coordinate model, provide a compelling characterization of the oscillons observed in this work. They emerge as large-amplitude, long-lived configurations in which the sub-kinks synchronize with remarkable quasi-periodicity.

As discussed, incorporating linear radiation effects into the collective-coordinate framework is crucial for reproducing many of the oscillon properties observed in the full-field dynamics. However, the decay of the oscillon, as well as the observed $K_n-{\bar K}_n$ annihilations, appears to be associated with a novel staccato-like mechanism. In this scenario, the central bion that binds the oscillon is partially confined by the outer sub-kinks, causing its dominant oscillation frequency to repeatedly cross the mass threshold. These threshold crossings generate successive bursts of radiation that progressively drain energy from the configuration and eventually trigger the abrupt collapse of the oscillon.

Overall, the results presented here outline a rich scenario that merits further exploration. They may provide insights into the emergence of coherent structures in physical systems \cite{Scott2003,Xu2023}, and they also suggest that extending the non-degenerate double sine-Gordon model, or similar constructions, to higher spatial dimensions, possibly including gauge fields, is a promising topic for future research.

\appendix
\section{Small oscillations  approximation}\label{Appen}

 As mentioned in Sec.~\ref{KnAKnsec}, after an initial transient stage, the oscillons evolve into a stable regime. Although the oscillation amplitude of the central binding bion is not necessarily small, the remaining sub-kink components oscillate coherently around their equilibrium separations. It is therefore reasonable to investigate the information provided by the small-oscillation approximation of the collective-coordinate equations of motion.

For the case $n=4$, we consider the coordinate vector

 \begin{equation}
 {\mathbf  x} (t)  = \left( x_{-2} \, , \,  x_{-1} \, , \,  x_{-2} \, , \,  x_{1} \, , \,   x_{2}  \right) \, e^{-i \omega \, t}. 
\end{equation}
 Linearizing eqs.~(\ref{ccm3}) around the equilibrium configuration leads to a eigenvalue problem, whose normal-mode frequencies are obtained from the eigenvalues of the following matrix:

\begin{equation}\label{wfreq1}
 \Upsilon_4 \,  = \begin{pmatrix}
  \, \frac{\varkappa_{2{\bar 2}} + \kappa_{12}}{M_b}  &  - \frac{ \kappa_{12} }{ \sqrt{M_a \, M_b}}  & 0  & 0\\  \\
- \frac{ \kappa_{12} }{ \sqrt{M_a \, M_b}}   & \frac{\kappa_{12} + \kappa_{0}}{M_a}  & \frac{ \kappa_{0}}{ \sqrt{M_a M_b}} & 0  \\ \\
0 & \frac{ \kappa_{0}}{ \sqrt{M_a M_b}}  & \frac{\kappa_{12} + \kappa_{0}}{M_a} & \frac{\kappa_{12} }{\sqrt{M_a M_b}}   \\ \\
0 & 0 &  \frac{\kappa_{12} }{\sqrt{M_a M_b}}  & \frac{\kappa_{12} \big(1 +i \, \frac{\pi^2 \beta_2(t_s)}{4 M_b} \omega \big)  }{ M_b}
\end{pmatrix} \, . 
 \end{equation}
 Here, $t_s$ represents a sufficiently large time for which the oscillon has already reached its stable phase. The parameters $M_a=M_{k_1}$ and $M_b=M_{k_2}$ are the sub-kink masses defined in eq.~(\ref{MkiSG}). The interaction strengths are characterized by the coupling constants $\varkappa_{2\bar{2}}$, associated with the $K_4-\bar{K}4$ interaction of eq.~(\ref{wKAK}), and   $ \kappa_{1,2}$  to  the contiguous  $k_i$ and $ k_{i+1}$  sub-kinks interactions eq.~(\ref{w1p1}). 
    
In eq.~(\ref{wfreq1}), damping is included only for the external sub-kink ${\bar k}_{2}$. The corresponding eigenvalues $\omega^2$ are readily obtained using an iterative approach. First, the damping is switched off by setting $\beta_2=0$, and the resulting undamped system is solved to determine the real normal-mode frequencies. These frequencies are then substituted into eq.~(\ref{wfreq1}), retaining the damping term, which yields complex eigenfrequencies $\omega=\omega_r+i\Gamma$. The imaginary part $\Gamma$ characterizes the decay rate of the corresponding normal mode.

 As an example, let us consider the parameter values $n=4$ and $\alpha=0.05$. The eigenfrequencies obtained from the matrix $\Upsilon_4$ are  $\omega  = \{ \, 0.0277 \, , 0.08   \, , 0.13   \, , 0.22 \, \}$.
 The lowest eigenfrequency lies below the radiation threshold $m_2=0.039$ and is therefore expected to remain stable against linear radiative losses. In contrast, the remaining three eigenfrequencies exceed this threshold and consequently contribute to the radiation emitted by the system. Although they initially remain below the second mass threshold, $m_1=0.71$, their frequencies increase during the evolution. The eventual crossing of this threshold gives rise to the staccato-like radiation mechanism.

 The observation that the external sub-kinks of the oscillon evolve in a highly synchronized manner suggests a further simplification of the collective-coordinate description. For arbitrary values of $n$, the oscillon may be approximated as a two-component system. The first component is the central bion, of mass $2 \, M_{k_\frac{n}{2}}$ , which remains bound to the rest of the configuration through the interaction force given in eq.~(\ref{w1p1}). The second component consists of the remaining external sub-kinks, treated as an effective degree of freedom with mass  $M_R=M_{K_{(n,\alpha)}}  -M_{k_\frac{n}{2}}$, where $M_{K_{(n,\alpha)}}$
 is the total multi-kink mass defined in eq.~(\ref{MKSM2}).

Within this approximation, the normal-mode frequencies are determined by the eigenvalues of the following matrix:

\begin{equation}\label{wfreq2}
\Upsilon_2 =\begin{pmatrix}
\frac{\varkappa_{2{\bar 2}} + \kappa_{12}}{M_{k_\frac{n}{2}}}  &  \,\,\,\,\, - \frac{ \kappa_{12} }{ \sqrt{M_R \, M_{k_\frac{n}{2}}}} \\  \\
  \frac{\kappa_{12} }{\sqrt{M_R M_{k_\frac{n}{2}}}}  & \,\,\,\,\, \frac{\kappa_{12} \big(1 +i \, \frac{\pi^2 \beta_2(t_s)}{4 M_R} \omega \big)  }{ M_R}
\end{pmatrix} 
 \end{equation}
  Here, for simplicity, we define $ \kappa_{12}   \equiv  \kappa_{{(\frac{n}{2}-1)  -  \frac{n}{2}}}$.  
Although the exact analytical expressions for the eigenvalues can be obtained explicitly, a simpler and physically transparent result follows from expanding in powers of the small parameter $M_{k_{\frac{n}{2}}}/M_R$. Since this ratio is typically small, such an approximation is well justified. A straightforward calculation then yields the following simplified expressions for the two normal-mode frequencies.

The low-frequency mode is given by

\begin{align}\label{wL}
\omega_L^2 \, =  \,  \frac{ \alpha \, m^3_ {(\frac{n}{2} -1  ) }}{M_R}  & \, f_c  \,  \approx \frac{ \pi \,  \alpha }{16 \, n} \, f_c, 
\quad \quad \quad T_L = \frac{2 \, \pi}{\omega_L} \approx  8 \sqrt{ \frac{ \pi \, n}{f_c \, \alpha}},  \\ \nonumber 
  f_c &=
  \left[1+  \frac{\alpha \,  m^3_ {(\frac{n}{2} -1  )} }{3 \, C_{n, \alpha} \, \left( m_{\,  \frac{n}{2}-1} -m_{\frac{n}{2}}   \right) } \right]^{-1} , 
 \end{align}
Here, we have used the approximation that for $n\geq 4$, the effective mass satisfies Here, we have used the approximation that for $n\geq 4$, the effective mass satisfies.  $M_R \, \approx  M_{K_{(n,\alpha)}}  \approx  16 \, n / \pi$. 
as follows from eq.~(\ref{MKSM3}). We have also introduced the coefficient $f_c$, which depends only weakly on both $\alpha$ and $n$. The resulting low-frequency mode satisfies $\omega_L < m_{\frac{n}{2}}$, remaining below the radiation threshold. Consequently, this mode is unable to efficiently emit radiation and is associated with the long period $T_L$ that characterizes the oscillon envelope.

The second eigenvalue corresponds to the high-frequency mode,

\begin{equation}\label{wH}
\omega_H^2 \, =  \,  \frac{\alpha }{M_{k_{\frac{n}{2}}}} \, \left[    m^3_ {(\frac{n}{2} -1  )}  + C_{n, \alpha} \, \left( m_{\,  \frac{n}{2}-1} -m_{\frac{n}{2}}   \right)  \right]  , \quad  \Gamma = \frac{1}{2} \,    \frac{\alpha \,  m^3_ {(\frac{n}{2} -1  )}}{M^2_{k_{\frac{n}{2}}}} \,  \beta_2(t_s) \, . 
 \end{equation}

In general, this frequency lies above the radiation threshold. Consequently, the mode can couple to propagating radiation states, and its eigenfrequency acquires a nonzero imaginary part. The corresponding radiative linewidth, $\Gamma$, determines the rate at which energy is lost through the linear radiation channel. An estimate of the oscillon lifetime associated with this process can therefore be obtained as

 \begin{equation}\label{tau}
\tau_{osc} \, \approx    \frac{M^2_{k_{\frac{n}{2}}}} {\alpha \, \, \beta_2(t_s) \, \, m^3_ {(\frac{n}{2} -1  )}}.
\end{equation}
Considering again the $n=4$ and $\alpha=0.05$ example, we obtain $\omega_L  = 0.0277$ and  $\omega_H  = 0.22$
that  coincide with the smallest and highest of the  eigenvalues of eq.~(\ref{wfreq1}): $\omega  = \{ \, 0.0277 \, , 0.08   \, , 0.13   \, , 0.22 \, \}$. 
 
 

\bibliography{CnD2sG.bib} 

\end{document}